\documentclass[aip,amsmath,amssymb,reprint,citeautoscript]{revtex4-1}

\usepackage{graphicx}% Include figure files
\usepackage{bm}% bold math
\usepackage{color}
\usepackage{stmaryrd}

\newcommand{\diff}{\mathrm{d}}

\begin{document}

\preprint{AIP/JCP19-AR-04544}

%\preprint{}

%\title{Tolman length and rigidity constants from density functional theory} %Title of paper

%\title{An accurate method for computing the Tolman length and rigidity constants for pure fluids and mixtures with density functional theory}
\title{Tolman lengths and rigidity constants from free-energy functionals -- General expressions and comparison of theories}



\author{P. Rehner}
\email[]{rehner@itt.uni-stuttgart.de}
%\thanks{}
\affiliation{Institute of Thermodynamics and Thermal Process Engineering, University of Stuttgart, Pfaffenwaldring 9, 70569 Stuttgart, Germany}
\author{A. Aasen}
\affiliation{Department of Energy and Process Engineering, Norwegian University of Science and Technology, NO-7491 Trondheim, Norway}
\affiliation{SINTEF Energy Research, NO-7465 Trondheim, Norway}
%\author{J. Gross}
%\email[]{gross@itt.uni-stuttgart.de}
%\affiliation{Institute of Thermodynamics and Thermal Process Engineering, University of Stuttgart, Pfaffenwaldring 9, 70569 Stuttgart, Germany}
\author{\O. Wilhelmsen}
\affiliation{Department of Energy and Process Engineering, Norwegian University of Science and Technology, NO-7491 Trondheim, Norway}
\affiliation{SINTEF Energy Research, NO-7465 Trondheim, Norway}


\date{\today}

\begin{abstract}
The leading order terms in a curvature expansion of the surface
tension, the Tolman length (first order), and rigidities (second
order) have been shown to play an important role in the description of
nucleation processes.
This work presents general and rigorous expressions to compute these quantities for any non-local density functional theory (DFT). The expressions hold for pure fluids and mixtures, and reduce to the known expressions from density gradient theory (DGT).
% In this work, we present a rigorous method for
% computing them to a high accuracy with non-local density functional
% theory (DFT) combined with the PCP-SAFT equation of state. General
% expressions that depend on the first order curvature expansion of the
% densities are provided for pure fluids and mixtures.
The framework is applied to a Helmholtz energy functional based on the perturbed chain polar statistical associating fluid theory (PCP-SAFT) and is used for an extensive investigation of curvature corrections for pure fluids and mixtures.
% The method is used to obtain curvature corrections for a range of fluids with full
% DFT in combination with the PCP-SAFT equation of state. 
Predictions from the full DFT are compared to two simpler theories: predictive
density gradient theory (pDGT), that has a density and temperature
dependent influence matrix derived from DFT, and DGT, where the influence parameter reproduces the surface tension as predicted from DFT. 
All models are based on the same equation of state and predict similar Tolman lengths and spherical rigidities for small molecules, but the deviations between DFT and DGT increase with chain length for the alkanes.
% We find that the models, which all rely
% on PCP-SAFT, give similar Tolman lengths ($\pm$~10\%) and spherical
% rigidities ($\pm$ 15\%). However, the deviation between the
% predictions from full DFT and DGT increases with chain length for the
% alkanes. 
% This is due to neglecting the chain contribution in DGT.  
For all components except water, we find that DGT underpredicts the value of the Tolman length, but overpredicts the value of the spherical rigidity. An important basis for the calculation is an accurate prediction of the planar surface tension. Therefore, further work is required to accurately extract Tolman lengths and rigidities of alkanols, because DFT with PCP-SAFT does not accurately predict surface tensions of these fluids.
\end{abstract}


\maketitle %\maketitle must follow title, authors, abstract and \pacs


\section{Introduction}
% This paragraph introduces the concept of curvature-dependent surface tension
% and some applications
The dependence of the surface tension on the interfacial curvature has
been discussed intensely over the last decades. One of the motivations
for studying the subject was the discrepancies between experiments and
theoretical predictions for nucleation rates in condensation and
evaporation~\cite{Iland2007,Vehkamaeki2006}. In classical nucleation
theory, the nucleation rate depends exponentially on the formation
energy of the nano-sized critical cluster. Therefore, it has been
hypothesized that omitting the curvature dependence of the surface
tension is the cause of the large discrepancies between theory and
experiments~\cite{Vehkamaeki2006,Wolde1998a,Wilhelmsen2015a,Nguyen2018}. The
curvature dependence of the surface tension also has implications for other important examples such as the properties of biomembranes\cite{Helfrich1973}, and wetting at the
nanoscale\cite{Kim2018}.

% This paragraph introduces the Helfrich expansion as a way to describe the
% curvature dependence
The first quantitative description was proposed by Tolman\cite{Tolman1949}. By
introducing the distance between the equimolar radius $R_e$ and the radius of
the surface of tension $R_s$, referred to as $\delta_T(R_s)$, he proposed the
expression
\begin{equation}
\sigma(R_s)=\frac{\sigma_0}{1+\frac{2\delta_T(R_s)}{R_s}}, % we should define Rs and Rt if we use them
\label{eq:Tolman}
\end{equation}
for the curvature dependent surface tension $\sigma(R_s)$ in relation to the
surface tension of a planar interface $\sigma_0$. In Eq.~\eqref{eq:Tolman}, the
curvature dependent Tolman length, $\delta_T(R_s)$ is often replaced by the
Tolman length of the planar interface $\delta$. In later
works~\cite{Wilhelmsen2015a,Wilhelmsen2015,Bruot2016,Aasen2018,Rehner2018}, it
was shown that with this approximation, Eq.~\eqref{eq:Tolman} is incapable of
representing the surface tension of small droplets. Instead, the second-order
expression by Helfrich\cite{Helfrich1973} has been established as the preferred
model to capture the curvature dependence of the surface tension. For an
arbitrarily curved interface, it reads
\begin{equation}
\sigma(J,K)=\sigma_0-\delta\sigma_0J+\frac{k}{2}J^2+\bar{k}K+\ldots,
\label{eq:Helfrich}
\end{equation}
where $\delta$ is the the Tolman length, $k$ is the bending rigidity and
$\bar{k}$ is the Gaussian rigidity. The interface is characterized locally by
the total curvature $J=1/R_1+1/R_2$ and the Gaussian curvature
$K=1/\left(R_1R_2\right)$, with $R_1$ and $R_2$ being the two principal radii. The expansion truncated at
second order is known as the Helfrich expansion, and the coefficients
$\sigma_0$, $\delta$, $k$ and $\bar k$ are referred to as Helfrich coefficients.
For spherical (index $s$) and cylindrical (index $c$) geometries,
Eq.~\eqref{eq:Helfrich} simplifies to
\begin{eqnarray}
\sigma^s(R)&&=\sigma_0-\frac{2\delta\sigma_0}{R}+\frac{2k+\bar{k}}{R^2}+\ldots~~~~~~\text{and}\label{eq:s}\\
\sigma^c(R)&&=\sigma_0-\frac{\delta\sigma_0}{R}+\frac{k}{2R^2}+\ldots
\label{eq:c}\end{eqnarray}
where $R$ is an arbitrarily chosen dividing surface. 
% This paragraph discusses estimates for particular fluids
The Tolman length for the vapor--liquid interface has been the subject of many
discussions and controversies, in particular its sign, since different routes to
obtain it have yielded different results. Due to its simplicity, most works in
the literature have considered the truncated and shifted Lennard--Jones (LJ)
fluid. Theoretical calculations based on free-energy functionals of the density
profile, such as density gradient theory (DGT) and non-local density functional
theory (DFT), have consistently given zero or negative
values~\cite{Blokhuis2006,Li2007,Block2010,Troester2012}. However, positive
values have been reported from Monte Carlo and molecular dynamics (MD)
simulations that compute the Tolman length from the pressure tensor
(see~\cite{Giessen2002,Lei2005,Horsch2012} and references therein). MD
simulations by van Giessen and Blokhuis, and also by Block et al.~have only
recently resulted in negative Tolman lengths around $-0.1$ in units of the LJ
diameter~\cite{Giessen2009,Sampayo2010,Block2010}; a consensus on this value is
emerging in the scientific community~\cite{Blokhuis2013}. Curvature corrections
for water have also been investigated
intensely\cite{Wilhelmsen2015a,Menzl2016,Joswiak2013,Joswiak2016,Kanduc2017,Kim2018,Leong2018}, also with
conflicting results on the sign of the Tolman length. Still, DGT and DFT have
been shown to agree \textit{quantitatively} with the predictions of recent
simulation studies for both the LJ
fluid\cite{Giessen2009,Blokhuis2013,Wilhelmsen2015} and
water\cite{Wilhelmsen2015a,Joswiak2013,Joswiak2016,Kanduc2017}, giving credibility to DFT and DGT as
methodologies to calculate curvature corrections.

% This paragraph discusses calculations from free-energy functionals
Free-energy functionals allow direct calculation of the curvature dependence of the
surface tension. In principle, the unknown coefficients in Eqs.~\eqref{eq:s} and
\eqref{eq:c} can be estimated by fitting a second order polynomial of the surface
tension as a function of the curvature, $1/R$~\cite{Rehner2018}. Since the surface tension of
large droplets and bubbles is very similar to $\sigma_0$, coefficients computed
in this manner have limited accuracy. A more accurate and rigorous route is to directly calculate the derivatives of the surface tension with
respect to curvature from the free-energy functional~\cite{Fisher1984}. This involves solving for
the first-order curvature expansion of the density profiles. The methodology was first presented
by Blokhuis and Bedeaux for pure fluids described by DGT~\cite{Blokhuis1993},
and later extended to mixtures described by DGT by Aasen et
al~\cite{Aasen2018}. Estimates of the Helfrich coefficients have also been
computed from various other free-energy
functionals\cite{Giessen1998,Barrett2009,Blokhuis2013,Rehner2018}. Still, a robust
method for calculating these coefficients rigorously from arbitrary free-energy functionals  and a systematic comparison of the values derived from different functionals for a range of fluids is missing.

%Eqs.~\eqref{eq:s} and \eqref{eq:c} show that $\delta$, $k$ and $\bar{k}$ can be determined by independently evaluating the curvature dependence of the surface tension in spherical and cylindrical geometries. By evaluating the surface tension of droplets and bubbles of different radii, the unknown coefficients can be estimated by fitting a second order polynomial in $1/R$. However, since Eq.~\eqref{eq:Helfrich} originates in a Taylor expansion around the planar interface, the expressions should be evaluated in the limit of infinitely large droplet or bubble radii~\cite{Fisher1984,Blokhuis1993}. Since the surface tension of large droplets and bubbles is very similar to $\sigma_0$, coefficients computed in this manner have limited accuracy. An alternative and more accurate route in DGT and DFT is to solve for the first-order function in the curvature expansion of the density profiles. The Tolman length and rigidities can then be computed by use of simple formulae. This was first done by Blokhuis and Bedeaux for pure fluids described by DGT~\cite{Blokhuis1993}, and later extended to mixtures by Aasen et al.~\cite{Aasen2018} In this work we present an extension of this methodology to full DFT combined with the PCP-SAFT equation of state, for pure fluids and mixtures.

% This paragraph explains what we do in THIS work
This work presents general expressions to compute the Tolman length
and rigidity constants for arbitrary free-energy functionals, that hold for
pure fluids and mixtures. These expressions are next applied to predict the
Helfrich coefficients to state-of-the-art accuracy for a range of pure fluids
and mixtures, using a non-local DFT based on the perturbed-chain polar statistical associating fluid theory (PCP-SAFT) equation of state. We present the first systematic comparison of these coefficients to
predictions from DGT and predictive density gradient
theory (pDGT)~\cite{Rehner2018a}. 
The comparison yields insight into the limits of using gradient theories for the description of curved interfaces. We will also shed light on the
impact of the underlying equation of state in DGT and non-local DFT.

% This paragraph outlines the paper
In section~\ref{sec:theory} we develop the general expressions for the Helfrich
coefficients, valid for any free-energy functional. In sections~\ref{sec:pure} and \ref{sec:mix}, coefficients for a range of pure fluids and mixtures are compared. In section~\ref{sec:conclusion}, we offer some concluding remarks.

\section{Theory \label{sec:theory}}
In this section we first review the general model-independent relations appearing in the curvature expansion. Subsequently we present the new expressions for the Helfrich coefficients for non-local DFT and shed light on how to treat the density dependence of the influence parameter in predictive density gradient theory. In all expressions, we consider isothermal conditions and paths. Bold symbols denote vector properties with respect to the components in the system.

\subsection{General relations for curvature expansion}
The aim of a curvature expansion is to determine thermodynamic properties of a curved interface by Taylor expanding around the planar interface. Therefore, every property $X$ that depends on the curvature is written as
\begin{equation}
X=X_0+\frac{X_1}{R}+\frac{X_2}{R^2}+\ldots,
\label{eq:ce_general}
\end{equation}
where the coefficients $X_i$ do not depend on the curvature. To be able to describe an arbitrarily shaped interface using the Helfrich expansion, the curvature expansion has to be performed in spherical and cylindrical coordinates. In the following, we derive expressions that are valid for both geometries, captured by the geometry factor $g$, which is 0 for a planar interface, 1 for a cylindrical interface and 2 for a spherical interface. Before presenting model-specific expressions, we derive general relations between the different properties of curved interfaces.

\subsubsection*{Gibbs--Duhem equation}
The bulk pressures in the liquid ($L$) and the vapor ($V$) phase are related to the chemical potential $\bm{\mu}$ and the density $\bm{\rho}$ of the system via the Gibbs-Duhem equation
\begin{align}
\bm{\rho}^V\cdot\diff\bm{\mu}=\diff p^V&&\text{and}&&\bm{\rho}^L\cdot\diff\bm{\mu}=\diff p^L.
\end{align}
Thus, the pressure difference $\Delta p=p^L-p^V$ is linked to the difference in densities $\Delta\bm{\rho}=\bm{\rho}^L-\bm{\rho}^V$ via
\begin{equation}
\Delta\bm{\rho}\cdot\diff\bm{\mu}=\diff\Delta p.
\end{equation}
Using Eq. \eqref{eq:ce_general} for all properties leads to the expression
\begin{multline}
\left(\Delta\bm{\rho}_0+\frac{\Delta\bm{\rho}_1}{R}+\ldots\right)\cdot\left(\bm{\mu}_1+\frac{2\bm{\mu}_2}{R}+\ldots\right)\diff\left(\frac{1}{R}\right)=\\
\left(\Delta p_1+\frac{2\Delta p_2}{R}+\ldots\right)\diff\left(\frac{1}{R}\right).
\end{multline}
Collecting terms with the same power of $R$ results in the relations
\begin{align}
\Delta p_1&=\Delta\bm{\rho}_0\cdot\bm{\mu}_1~~~~~~~~~~\text{and}\label{eq:GibbsDuhem1}\\
\Delta p_2&=\Delta\bm{\rho}_0\cdot\bm{\mu}_2+\frac{1}{2}\Delta\bm{\rho}_1\cdot\bm{\mu}_1.\label{eq:GibbsDuhem2}
\end{align}

\subsubsection*{Adsorption}
The adsorption, $\bm{\Gamma}$ refers to the amount of particles accumulated at the interface per surface area, and isdefined as
\begin{equation}
\bm{\Gamma}=\int\bm{\rho}^E(r)\left(\frac{r}{R}\right)^g\diff r,
\end{equation}
where we introduce the excess density
\begin{equation}
\bm{\rho}^E(r)=\bm{\rho}(r)-\bm{\rho}^L\Theta(R-r)-\bm{\rho}^V\Theta(r-R),
\end{equation}
with the Heaviside step function $\Theta(r)$. By changing the integration variable to $z=r-R$ and again collecting terms of the same order in curvature, the following expansion coefficients
\begin{align}
\bm{\Gamma}_0&=\int\bm{\rho}_0^E(z)\diff z,~~~~~~~~~~\text{and}\\
\bm{\Gamma}_1&=\int\left(\bm{\rho}_1^E(z)+gz\bm{\rho}_0^E(z)\right)\diff z
\end{align}
are obtained.

\subsubsection*{Gibbs adsorption equation}
The Gibbs adsorption equation
\begin{equation}
\diff\sigma=-\bm{\Gamma}\cdot\diff\bm{\mu}+\left[\frac{\partial\sigma}{\partial R}\right]_{T,\bm{\mu}}\diff R
\label{eq:GibbsAdsorption}
\end{equation}
links the adsorption to the surface tension $\sigma$. As we have not yet made a choice of dividing surface, the notional derivative, $\left[\frac{\partial\sigma}{\partial R}\right]_{T,\bm{\mu}}$ appears in the equation. The notial derivative describes the change in surface tension due to a change in the dividing surface, while keeping the physical system unaltered. The notional derivative also enters the general form of the Young-Laplace equation, as
\begin{equation}
\Delta p=\frac{g\sigma}{R}+\left[\frac{\partial\sigma}{\partial R}\right]_{T,\bm{\mu}}.
\label{eq:YoungLaplace}
\end{equation}
Substituting the notional derivative from Eq. \eqref{eq:YoungLaplace} into \eqref{eq:GibbsAdsorption} and expanding the resulting expression gives a general relation between the coefficients of the pressure difference and the surface tension, as
\begin{align}
\Delta p_0&=0,~~~~\Delta p_1=g\sigma_0~~~~\text{and}\nonumber\\
\Delta p_2&=-\bm{\Gamma}_0\cdot\bm{\mu}_1+(g-1)\sigma_1.
\label{eq:deltaP}
\end{align}

\subsubsection*{Density profiles}
The density profile of an open system is obtained as a stationary point of the grand potential functional $\Omega$. Using a Legendre transform, the equilibrium condition can be formulated in terms of the functional derivative of the Helmholtz energy, $F$ instead
\begin{equation}
\left.\frac{\delta\Omega}{\delta\bm{\rho}(\mathbf{r})}\right|_{T,V,\bm{\mu}}=0~~\Leftrightarrow~~\left.\frac{\delta F}{\delta\bm{\rho}(\mathbf{r})}\right|_{T,V}=\bm{\mu}.
\label{eq:EulerLagrange}
\end{equation}
A curvature expansion of Eq. \eqref{eq:EulerLagrange} gives
\begin{align}
\bm{\mu}_0&=\left(\frac{\delta F}{\delta\bm{\rho}(\mathbf{r})}\right)_0,~~\text{and}\\
\bm{\mu}_1&=\int\left(\frac{\delta^2F}{\delta\bm{\rho}(\mathbf{r})\delta\bm{\rho}(\mathbf{r}')}\right)_0\bm{\rho}_1(\mathbf{r}')\diff\mathbf{r}'.
\label{eq:EulerLagrange1_general}
\end{align}
The density profile of the planar interface and the first-order term in the curvature expansion of the density can be obtained by solving the above equations. It is not obvious from the general formulation how they should be solved. However, one important property can be derived. For free-energy functionals $F$, the expression
\begin{equation}
\int\left(\frac{\delta^2F}{\delta\bm{\rho}(\mathbf{r})\delta\bm{\rho}(\mathbf{r}')}\right)_0\nabla\bm{\rho}_0(\mathbf{r}')\diff\mathbf{r}'=\nabla\left(\frac{\delta F}{\delta\bm{\rho}(\mathbf{r})}\right)_0=0,
\end{equation}
vanishes at equilibrium. It follows, that if $\bm{\rho}_1(\mathbf{r})$ is a solution of Eq. \eqref{eq:EulerLagrange1_general}, $\bm{\rho}_1(\mathbf{r})+\varepsilon\nabla\bm{\rho}_0(\mathbf{r})$ is also a solution for any value of $\varepsilon$.

\subsubsection*{Surface tension}
The surface tension is defined as the excess grand potential per surface area, $\sigma=\frac{\Omega^E}{A}$. Using the geometry factor $g$, it can be expressed as
\begin{equation}
\sigma=\int\left(f-\bm{\rho}\cdot\bm{\mu}+p^{LV}\right)\left(\frac{r}{R}\right)^g\diff r
\end{equation}
with the Helmholtz energy density, $f$ and the pressure of the bulk phases, $p^{LV}=p^L\Theta(r-R)+p^V\Theta(R-r)$, which is related to the adsorption via the Gibbs-Duhem equation. After a few simplification steps and identifying the excess grand potential density of the planar interface, $\Delta\omega_0=f_0-\bm{\rho}_0\cdot\bm{\mu}_0+p_0$, the resulting expressions for the coefficients are
\begin{align}
\sigma_0=&\int\Delta\omega_0\diff z\\
\sigma_1=&\int\left(f_1-\bm{\rho}_1\cdot\bm{\mu}_0\right)\diff z+g\int\Delta\omega_0z\diff z-\bm{\mu}_1\cdot\bm{\Gamma}_0
\label{eq:gamma1}\\
\sigma_2=&\int\left(f_2-\bm{\rho}_2\cdot\bm{\mu}_0-\frac{1}{2}\bm{\rho}_1\cdot\bm{\mu}_1\right)\diff z\nonumber\\
&+g\int\left(f_1-\bm{\rho}_1\cdot\bm{\mu}_0\right)z\diff z+\frac{g(g-1)}{2}\int\Delta\omega_0z^2\diff z\nonumber\\
&-\frac{g}{2}\bm{\mu}_1\cdot\int\bm{\rho}_0^Ez\diff z-\bm{\mu}_2\cdot\bm{\Gamma}_0-\frac{1}{2}\bm{\mu}_1\cdot\bm{\Gamma}_1
\label{eq:gamma2}
\end{align}
It is tempting to neglect the first term in both the first and second order expressions for the surface tension as we find
\begin{multline}
\int\left(f_1-\bm{\rho}_1\cdot\bm{\mu}_0\right)\diff\mathbf{r}=\\
\iint\left(\left(\frac{\delta f(\mathbf{r})}{\delta\bm{\rho}(\mathbf{r}')}\right)_0\cdot\bm{\rho}_1(\mathbf{r}')-\left(\frac{\delta f(\mathbf{r}')}{\delta\bm{\rho}(\mathbf{r})}\right)_0\cdot\bm{\rho}_1(\mathbf{r})\right)\diff\mathbf{r}\diff\mathbf{r}'
\end{multline}
which is strictly zero. However, in Eq. \eqref{eq:gamma1}, the integration is over $z$. We have to take into account that the integration takes place in a curvilinear coordinate system, even if one is interested in the limit of zero curvature.

\subsubsection*{Path through the metastable region}
Although the norms of the vectors $\bm{\mu}_1$ and $\bm{\mu}_2$ are fixed by Eqs. \eqref{eq:GibbsDuhem1}, \eqref{eq:GibbsDuhem2} and \eqref{eq:deltaP}, their directions represent degrees of freedom. Every point in the metastable region is defined by its temperature and chemical potentials. However, there is an infinite number of possible starting points on the phase envelope and paths towards the metastable point, which are each equipped with their own expansion coefficients. In a previous study\cite{Aasen2018}, it was shown that a straight path (i.e. $\bm{\mu}_2\propto\bm{\mu}_1$) where the composition of the liquid phase $\bm{x}^L$ is kept constant in the first order term gives low errors in the expansion. To obtain the coefficients of the chemical potential for this choice of path, the first order coefficient of the total liquid density is calculated as
\begin{equation}
\rho^L_1=\frac{g\sigma_0}{\left(\bm{x}^L\right)^T\left(\bm{\mu}_\rho^L\right)_0\Delta\bm{\rho}_0},
\end{equation}
from which the chemical potential,
\begin{equation}
\bm{\mu}_1=\rho_1^L\left(\bm{\mu}_\rho^L\right)_0\bm{x}^L
\end{equation}
and the vapor partial densities,
\begin{equation}
\bm{\rho}_1^V=\left(\bm{\mu}_\rho^V\right)_0^{-1}\bm{\mu}_1
\end{equation}
follow. The second order coefficient for the chemical potential can be derived from Eqs. \eqref{eq:GibbsDuhem2} and \eqref{eq:deltaP} as
\begin{equation}
\bm{\mu}_2=\frac{(g-1)\sigma_1-\left(\bm{\Gamma}_0+\frac{1}{2}\Delta\bm{\rho}_1\right)\cdot\bm{\mu}_1}{g\sigma_0}\bm{\mu}_1.
\end{equation}

\subsubsection*{Choice of dividing surface}
The equations derived so far are valid for any choice of dividing surface. However, to be able to evaluate the expressions, a choice has to be made. The first option is the surface of tension $R_s$, for which the notional derivative of the surface tension vanishes. Thus, the usual form of the Young Laplace equation $\Delta p=\frac{g\sigma}{R_s}$ is valid and the Gibbs adsorption equation simplifies to
\begin{align}
\sigma_1=-\bm{\mu}_1\cdot\bm{\Gamma}_0&&\text{and}&&\sigma_2=-\bm{\mu}_2\cdot\bm{\Gamma}_0-\frac{1}{2}\bm{\mu}_1\cdot\bm{\Gamma}_1.
\end{align}
The second important option is the equimolar dividing surface or its generalization to multicomponent mixtures, the Koenig surface\cite{Koenig1950} $R_k$, which is defined by $\bm{\Gamma}\cdot\diff\bm{\mu}=0$. As opposed to the surface of tension which is a state function, the Koenig surface is path dependent. For specific applications, one choice might be superior. However, it is important to keep in mind that neither the planar surface tension nor the Tolman length depend on the dividing surface and there are simple model-independent relations for the rigidity constants for different dividing surfaces\cite{Aasen2018}.

In the density profile of the planar interface, the dividing surface is fixed by finding any density profile that solves the zeroth order Euler--Lagrange equation and then shifting the $z$-axis by the position of the Koenig surface $z_{k0}$ or the surface of tension $z_{s0}$. The values for the two surfaces are given by
\begin{align}
z_{k0}=\frac{\bm{\mu}_1\cdot\bm{\Gamma}_0}{g\sigma_0}&&\text{and}&&z_{s0}=\frac{\sigma_1+\bm{\mu}_1\cdot\bm{\Gamma}_0}{g\sigma_0}.
\end{align}
From these relations, we obtain an expression for the Tolman length
\begin{equation}
z_{k0}-z_{s0}=\frac{-\sigma_1}{g\sigma_0}=\delta.
\end{equation}
With a similar procedure, the correct solution of the first order Euler--Lagrange equation is found by first finding any solution $\bm{\tilde{\rho}}_1(z)$ and then obtaining the actual solution as $\bm{\rho}_1(z)=\bm{\tilde{\rho}}_1(z)+\varepsilon\bm{\rho}_0'(z)$ with
\begin{equation}
\varepsilon=\frac{2\bm{\mu}_2\cdot\bm{\Gamma}_0+\bm{\mu}_1\cdot\bm{\tilde{\Gamma}}_1}{g\sigma_0}
\label{eq:epsilon_k}
\end{equation}
for the Koenig surface and
\begin{equation}
\varepsilon=\frac{\sigma_2+2\bm{\mu}_2\cdot\bm{\Gamma}_0+\bm{\mu}_1\cdot\bm{\tilde{\Gamma}}_1}{g\sigma_0}
\label{eq:epsilon_s}
\end{equation}
for the surface of tension.

\subsection{Non-local density functional theory\label{sec:DFT}}
In non-local density functional theory (DFT), the Helmholtz energy $F[\bm{\rho}(\mathbf{r})]=\int f[\bm{\rho}(\mathbf{r})]\diff\mathbf{r}$ and the Helmholtz energy density $f[\bm{\rho}(\mathbf{r})]$ are functionals of the density profiles $\bm{\rho}(\mathbf{r})$ of all components. In most DFT approaches, the Helmholtz energy density can be written as a function of any number of weighted densities $n_\alpha$. This includes functionals based on fundamental measure theory (FMT)\cite{Rosenfeld1989}, local and weighted density approximations\cite{Tarazona1985} and mean-field theory\cite{Blokhuis2013}. The weighted densities are obtained by convolving the density profile with corresponding weight functions, $\bm{\omega}_\alpha$ in three dimensions
\begin{equation}
n_\alpha(\mathbf{r})=\bm{\rho}\stackrel{3D}{\otimes}\bm{\omega}_\alpha=\int\bm{\rho}(\mathbf{r}-\mathbf{r}')\cdot\bm{\omega_\alpha}(\mathbf{r}')\diff\mathbf{r}',
\end{equation}
where the sum in the inner product is over all components. To calculate the expansion coefficients, the curvature expansion of the convolution integral is required. This is straightforward for a spherical geometry, but significantly more tedious in cylindrical coordinates, as shown in Appendix \ref{app:convolutions}. The zeroth and first order expressions for the weighted densities can be written using one dimensional convolution integrals,
\begin{align}
n_{\alpha 0}&=\bm{\rho}_0\otimes\bm{\omega}_\alpha~~~~~~~\text{and}
\label{eq:n_alpha_0}\\
n_{\alpha 1}&=\bm{\rho}_1\otimes\bm{\omega}_\alpha-\frac{g}{2}\bm{\rho}_0\otimes\left(z\bm{\omega}_\alpha\right).
\label{eq:n_alpha_1}
\end{align}
%The density profile is solved for by minimizing the grand potential $\Omega=F[\bm{\rho}(\mathbf{r})]-\bm{\mu}\cdot\bm{N}$ of the system, which leads to the condition for the equilibrium profile
%\begin{align}
%\left(\frac{\delta\Omega}{\delta\bm{\rho}(\mathbf{r})}\right)_{T,V,\bm{\mu}}=0&&\text{or}&&\frac{\delta F}{\delta\bm{\rho}(\mathbf{r})}=\bm{\mu}.
%\end{align}
For the curvature expansion, the first
\begin{align}
f_1&=\sum_\alpha f_{\alpha 0}n_{\alpha 1}~~~~\text{and second}
\label{eq:f_1}\\
f_2&=\frac{1}{2}\sum_\alpha f_{\alpha 1}n_{\alpha 1}+\sum_{\alpha}f_{\alpha 0}n_{\alpha 2}
\end{align}
order coefficients of the Helmholtz energy density are required.
Here,
\begin{align}
f_{\alpha 0}=\left(\frac{\partial f}{\partial n_\alpha}\right)_0&&\text{and}&&f_{\alpha\beta 0}=\left(\frac{\partial^2f}{\partial n_\alpha\partial n_\beta}\right)_0
\end{align}
is shorthand for the zeroth order first and second partial derivatives of the Helmholtz energy density and $f_{\alpha 1}=\sum_\beta f_{\alpha\beta 0}n_{\beta 1}$ is the corresponding first order expression. The same concept as for the weighted densities is used to obtain the first and second order expressions for the Euler--Lagrange equation, giving
\begin{align}
\bm{\mu}_0&=\left(\frac{\delta F}{\delta\bm{\rho}}\right)_0=\sum_\alpha f_{\alpha 0}\otimes\bm{\omega}_\alpha~~~~~~~\text{and}
\label{eq:EulerLagrange0}\\
\bm{\mu}_1&=\left(\frac{\delta F}{\delta\bm{\rho}}\right)_1=\sum_\alpha\left(f_{\alpha 1}\otimes\bm{\omega}_\alpha-\frac{g}{2}f_{\alpha 0}\otimes\left(z\bm{\omega}_\alpha\right)\right).
\label{eq:EulerLagrange1}
\end{align}
Using Eqs. \eqref{eq:n_alpha_1}, \eqref{eq:f_1} and \eqref{eq:EulerLagrange0}, the first term in the general expression \eqref{eq:gamma1} for $\sigma_1$ can be simplified as
\begin{align*}
\int&\left(f_1-\bm{\rho}_1\cdot\bm{\mu}_0\right)\diff z\\
&=\int\sum_\alpha\Big(f_{\alpha 0}\left(\bm{\rho}_1\otimes\bm{\omega}_\alpha\right)-\frac{g}{2}f_{\alpha 0}\left(\bm{\rho}_0\otimes\left(z\bm{\omega}_\alpha\right)\right)\\
&~~~~~~~~~~~~~~~~~~~~~~~~~~~~~~~~~~~~~~-\bm{\rho}_1\left(f_{\alpha 0}\otimes\bm{\omega}_\alpha\right)\Big)\diff z\\
&=-\frac{g}{2}\int\sum_\alpha f_{\alpha 0}\left(\bm{\rho}_0\otimes\left(z\bm{\omega}_\alpha\right)\right)\diff z.
\end{align*}
Through its definition $\sigma_1=-g\delta\sigma_0$, the Tolman length follows as
\begin{multline}
\delta\sigma_0=\frac{1}{2}\int\sum_\alpha f_{\alpha 0}\left(\bm{\rho}_0\otimes\left(z\bm{\omega}_\alpha\right)\right)\diff z\\
-\int\Delta\omega_0z\diff z+\frac{1}{2}\bm{\mu}_1^s\cdot\bm{\Gamma}_0.
\label{eq:Tolman_DFT}
\end{multline}
For different Helmholtz energy functionals, the Tolman length depends only on the planar density profile \cite{Blokhuis1993,Blokhuis2013}. In a similar albeit more elaborate fashion, the rigidity constants are obtained as
\begin{multline}
k=-\frac{1}{4}\int\sum_\alpha f_{\alpha 0}\left(\bm{\rho}_0\otimes\tilde{\bm{\omega}}_\alpha\right)\diff z\\
-\frac{1}{4}\int\sum_\alpha\left(\bm{\rho}_1^s\cdot\left(f_{\alpha 0}\otimes\left(z\bm{\omega}_\alpha\right)\right)+f_{\alpha 1}^s\left(\bm{\rho}_0\otimes\left(z\bm{\omega}_\alpha\right)\right)\right)\diff z\\
-\frac{1}{2}\bm{\mu}_1^s\cdot\int\bm{\rho}_0^Ez\diff z-2\bm{\mu}_2^c\cdot\bm{\Gamma}_0-\frac{1}{4}\bm{\mu}_1^s\cdot\bm{\Gamma}_1^s
\label{eq:k_DFT}
\end{multline}
and
\begin{multline}
\bar{k}=\int\Delta\omega_0z^2\diff z+\frac{1}{2}\int\sum_\alpha f_{\alpha 0}\left(\bm{\rho}_0\otimes\tilde{\bm{\omega}}_\alpha\right)\diff z\\
-\int\sum_\alpha f_{\alpha 0}\left(\bm{\rho}_0\otimes\left(z\bm{\omega}_\alpha\right)\right)z\diff z+\left(4\bm{\mu}_2^c-\bm{\mu}_2^s\right)\cdot\bm{\Gamma}_0.
\label{eq:kbar_DFT}
\end{multline}
The full derivation of these expressions is shown in the supplementary material. The gaussian rigidity, $\bar{k}$ also does not depend on $\bm{\rho}_1$, but the bending rigidity, $k$ does. To calculate all Helfrich coefficients, it is therefore necessary to calculate $\bm{\rho}_0$ and $\bm{\rho}_1$ from the zeroth and first order expressions of the Euler--Lagrange equation.
It is, however, only necessary to calculate $\rho_1$ for one geometry, as all first order expressions are proportional to the geometry factor $g$ and therefore $\bm{\rho}_1^s=2\bm{\rho}_1^c$, i.e. the value for a spherical geometry is twice the value for a cylindrical geometry. Further, if $\bm{\rho}_1$ is a solution to Eq. \eqref{eq:EulerLagrange1}, $\bm{\rho}_1+\varepsilon\bm{\rho}_0'$ is also a solution for any value of $\varepsilon$. A thorough investigation of Eq. \eqref{eq:k_DFT} reveals, however, that $k$  does not depend on the value of $\varepsilon$. Therefore, it is sufficient to find any solution of Eq. \eqref{eq:EulerLagrange1} to compute the Helfrich coefficients.
% OW: I think this will confuse people more than it helps, have seen no such applications are stated below in the literature, so it is not so relevant.
%However, if the goal is to use the density profiles $\bm{\rho}_1$ to predict the density profiles of droplets, a correct solution that satisfies the constraints for the chosen dividing surface must be determined by using Eq. \eqref{eq:epsilon_k} or \eqref{eq:epsilon_s}.

Although different numerical methods have been applied\cite{Mairhofer2017}, the standard method in DFT is to solve for the density profiles by use of fixed point iteration. To calculate the planar density profile, the functional derivative in Eq. \eqref{eq:EulerLagrange0} is split into an ideal gas contribution and a residual, resulting in the iteration
\begin{equation}
\bm{\rho}_0=\exp{\left(\frac{1}{k_BT}\left(\bm{\mu}_0-\left(\frac{\delta F^\mathrm
{res}}{\delta\bm{\rho}}\right)_0\right)\right)}.
\end{equation}
The same concept can be used to solve for the curvature correction, giving
\begin{equation}
\bm{\rho}_1=\frac{\bm{\rho}_0}{k_BT}\left(\bm{\mu}_1-\left(\frac{\delta F^\mathrm{res}}{\delta\bm{\rho}}\right)_1\right).
\end{equation}
The convergence of the iteration can be sped up significantly by using an Anderson mixing scheme\cite{Anderson1965,Mairhofer2017}.

\subsection{Predictive density gradient theory}
In predictive density gradient theory (pDGT)\cite{Rehner2018a}, the Helmholtz energy functional has the form
\begin{equation}
F[\bm{\rho}(\mathbf{r})]=\int\left(f^\text{eos}(\bm{\rho})+\frac{1}{2}\nabla\bm{\rho}^T\bm{C}(\bm{\rho})\nabla\bm{\rho}\right).
\end{equation}
The difference compared to standard density or square gradient theory comes from the density and temperature dependence of the influence matrix, $\bm{C}$. Both the influence matrix and the bulk Helmholtz energy density, $f^\text{eos}$ can be related to the Helmholtz energy density in non-local DFT as
\begin{align}
f^\text{eos}(\bm{\rho})&=f(\lbrace n_\alpha^b\rbrace)~~~~~~\text{and}\\
\mathbf{C}(\bm{\rho})&=-\sum_{\alpha\beta}f_{\alpha\beta}(\lbrace n_\alpha^b\rbrace)\left(\bm{\omega}_\alpha^0{\bm{\omega}_\beta^2}^T+\bm{\omega}_\alpha^2{\bm{\omega}_\beta^0}^T\right)\nonumber
\end{align}
with the moments of the weight functions
\begin{align}
\bm{\omega}_\alpha^0=4\pi\int\limits_0^\infty\bm{\omega}_\alpha(r)r^2\diff r&&\text{and}&&\bm{\omega}_\alpha^2=\frac{2\pi}{3}\int\limits_0^\infty\bm{\omega}_\alpha(r)r^4\diff r.\nonumber
\end{align}
and the weighted densities evaluated for local bulk conditions $n_\alpha^b=\bm{\rho}\cdot\bm{\omega}_\alpha^0$. The expressions for the Helfrich coefficients are the same as for standard DGT\cite{Blokhuis1993,Aasen2018}.
\begin{align}
\sigma_0=&\int{\bm{\rho}_0'}^T\bm{C}_0\bm{\rho}_0'\diff z\\
\delta\sigma_0=&-\int{\bm{\rho}_0'}^T\bm{C}_0\bm{\rho}_0'z\diff z+\frac{1}{2}\bm{\mu}_1^s\cdot\bm{\Gamma}_0\\
k=&-\frac{1}{2}\int{\bm{\rho}_0'}^T\bm{C}_0\bm{\rho}_1^s\diff z-\frac{1}{2}\bm{\mu}_1^s\cdot\int\bm{\rho}_0^Ez\diff z\nonumber\\
&-2\bm{\mu}_2^c\cdot\bm{\Gamma}_0-\frac{1}{4}\bm{\mu}_1^s\cdot\bm{\Gamma}_1^s\\
\bar{k}=&\int{\bm{\rho}_0'}^T\bm{C}_0\bm{\rho}_0'z^2\diff z+\left(4\bm{\mu}_2^c-\bm{\mu}_2^s\right)\cdot\bm{\Gamma}_0
\end{align}
For pDGT, the density dependence of the influence matrix has to be taken into account when calculating the density profile from the Euler--Lagrange equation. Therefore, we propose a slight modification to the approach previously suggested for a constant influence matrix\cite{Aasen2018}. Similar to the method for planar interfaces\cite{Rehner2018a}, we use the geometric combining rule $\bm{C}=\bm{c}\bm{c}^T$. The vector $\bm{c}$ contains the square root of the diagonal elements of the influence matrix. The advantage of this approach is that it leads to a separation of the Euler--Lagrange equation into a system of algebraic equations
\begin{equation}
f_{\bm{\rho}}^\mathrm{eos}-\bm{\mu}=\alpha\bm{c}
\label{eq:algebraic_equations}
\end{equation}
with the unknown $\alpha$ and one differential equation. To obtain it, we introduce $u=\bm{c}\cdot\bm{\rho}'$ and use it in the integrated form of the Euler--Lagrange equation
\begin{align}
\left(f^\mathrm{eos}-\bm{\rho}\cdot\bm{\mu}-\frac{1}{2}{\bm{\rho}'}^T\bm{C}\bm{\rho}'\right)'-\frac{g}{r}{\bm{\rho}'}^T\bm{C}\bm{\rho}'=0.
\label{eq:Noether}
\end{align}
The above equation is applicable to planar ($g=0$), cylindrical ($g=1$) and spherical ($g=2$) geometries. By identifying ${\bm{\rho}'}^T\bm{C}\bm{\rho}'=u^2$ and $\left(f^\mathrm{eos}-\bm{\rho}\cdot\bm{\mu}\right)'=\alpha u$, Eq. \eqref{eq:Noether} can be written compactly as
\begin{equation}
\left(f^\mathrm{eos}-\bm{\rho}\cdot\bm{\mu}-\frac{1}{2}u^2\right)'-\frac{g}{r}u^2=0.
\label{eq:Noether_u}
\end{equation}
or after evaluating the gradient and dividing by $u$ as
\begin{equation}
u'=\alpha-\frac{g}{r}u.
\end{equation}
To find the planar density profile $\bm{\rho}_0$, the system is discretized along a path function $s$, which has to be monotonous in the interface. Different choices for this path function have been proposed, including the density of the least volatile component\cite{Miqueu2004,Miqueu2005,Mairhofer2017a}, the so-called weighted molecular density $\frac{\bm{c}^T\bm{\rho}_0}{\sqrt{\bm{c}^T\bm{c}}}$ by Kou et al.~\cite{Kou2015}, or the unscaled version $\bm{c}^T\bm{\rho}_0$ of Liang et al.~\cite{Liang2016}. At every discretization point, the zeroth order expansion of Eq. \eqref{eq:algebraic_equations}
\begin{equation}
f_{\bm{\rho} 0}^\mathrm{eos}-\bm{\mu}_0=\alpha_0\bm{c}_0
\end{equation}
has to be solved. For the planar interface, Eq. \eqref{eq:Noether_u} can be integrated analytically to give $u_0$ as
\begin{equation}
u_0=\sqrt{2\left(f_0^\mathrm{eos}-\bm{\rho}_0\cdot\bm{\mu}_0+p_0\right)}=\sqrt{2\Delta\omega_0^\mathrm{eos}}
\end{equation}
where the pressure $p_0$ appears as a constant of the integration. Finally, using the zeroth order term of the definition of $u$, the $z$-axis is obtained as
\begin{equation}
u_0=\bm{c}_0\cdot\bm{\rho}_0'=\bm{c}_0\cdot\frac{\diff\bm{\rho}_0}{\diff s}\frac{\diff s}{\diff z}~~\Rightarrow~~z=\int\frac{\bm{c}_0\cdot\frac{\diff\bm{\rho}_0}{\diff s}}{\sqrt{2\Delta\omega_0^\mathrm{eos}}}\diff s
\end{equation}
The integration constant is determined by the choice of dividing surface, analogous to Sec.~\ref{sec:DFT}.

For the curvature correction, the first order expression of Eq. \eqref{eq:algebraic_equations}, which is the linear algebraic equation
\begin{equation}
\left(f_{\bm{\rho}\bm{\rho}0}^\mathrm{eos}-\alpha_0\bm{c}_{\bm{\rho} 0}\right)\bm{\rho}_1=\bm{\mu}_1+\alpha_1\bm{c}_0,
\end{equation}
has to be solved simultaneously with the linear differential equation
\begin{equation}
u_1'=\bm{c}_0\bm{\rho}_1''+\bm{c}_0'\bm{\rho}_1'+\bm{c}_1\bm{\rho}_0''+\bm{c}_1'\bm{\rho}_0'=\alpha_1-gu_0
\end{equation}
for the density profile $\bm{\rho}_1$ and $\alpha_1$. Again, the solution corresponding to a specific dividing surface can be found using Eq. \eqref{eq:epsilon_k} or Eq.~\eqref{eq:epsilon_s}.

%  For planar ($g=0$), cylindrical ($g=1$) and spherical ($g=2$) geometry it can be written as
% \begin{equation}
% f^\text{eos}_{\bm{\rho}}-\bm{\mu}+\bm{\Psi}_{\bm{\rho}}\bm{\rho}''+\bm{\Psi}''-\frac{g}{r}\bm{C}\bm{\rho}'=0
% \label{eq:EulerLagrangepDGT}
% \end{equation}
% where we use $\bm{\Psi}$ to avoid the calculation of derivatives of $\bm{C}$. It is defined as
% \begin{align}
% \bm{\Psi}&=\sum_\alpha f_\alpha(\lbrace n_\alpha^b\rbrace)\bm{\omega}_\alpha^2~~~~\text{and}\\
% \bm{\Psi}_{\bm{\rho}}&=\sum_{\alpha\beta}f_{\alpha\beta}(\lbrace n_\alpha^b\rbrace)\bm{\omega}_\alpha^2{\bm{\omega}_\beta^0}^T
% \end{align}
% and is connected to the influence matrix via $\bm{C}=-\bm{\Psi}_{\bm{\rho}}-\bm{\Psi}_{\bm{\rho}}^T$. To solve for $\bm{\rho}_0$ and $\bm{\rho}_1$ Eq. \eqref{eq:EulerLagrangepDGT} has to be curvature expanded, as
% \begin{equation}
% f^\text{eos}_{\bm{\rho}0}-\bm{\mu}_0+\bm{\Psi}_{\bm{\rho}0}\bm{\rho}_0''+\bm{\Psi}_0''=0
% \end{equation}
% and
% \begin{equation}
% f^\text{eos}_{\bm{\rho}\bm{\rho}0}\bm{\rho}_1-\bm{\mu}_1+\bm{\Psi}_{\bm{\rho}1}\bm{\rho}_0''+\bm{\Psi}_{\bm{\rho}0}\bm{\rho}_1''+\bm{\Psi}_1''-g\bm{C}_0\bm{\rho}_0'=0
% \end{equation}
% with
% \begin{align}
% \bm{\Psi}_1=\bm{\Psi}_{\bm{\rho}0}\bm{\rho}_1&&\text{and}&&\bm{\Psi}_{\bm{\rho}1}=\bm{\Psi}_{\bm{\rho}\bm{\rho}0}\bm{\rho}_1
% \label{eq:Psirho1}
% \end{align}

\subsection{The PCP-SAFT equation of state}
The expressions shown in Sec.~\ref{sec:DFT} are valid for any Helmholtz energy functional that can be written in terms of weighted densities. To calculate Helfrich coefficients for a variety of pure components and mixtures, we apply it to the Helmholtz energy functional based on the perturbed-chain polar statistical associating fluid theory (PCP-SAFT) equation of state\cite{Gross2001,Gross2005,Gross2006}. Similar to the equation of state, the residual Helmholtz energy functional is split into different contributions, each modeling different intermolecular interactions, as
\begin{multline}
F^\mathrm{res}[\bm{\rho}(\mathbf{r})]=F^\mathrm{hs}[\bm{\rho}(\mathbf{r})]+F^\mathrm{chain}[\bm{\rho}(\mathbf{r})]\\+F^\mathrm{assoc}[\bm{\rho}(\mathbf{r})]+F^\mathrm{att}[\bm{\rho}(\mathbf{r})].
\end{multline}
For the hard-sphere (hs) contribution, fundamental measure theory\cite{Rosenfeld1989,Roth2010} has been well established. We used the version proposed by Roth\cite{Roth2002} and Yu and Wu\cite{Yu2002} that uses vector weight functions. If those are to be avoided, the version by Kierlik and Rosinberg\cite{Kierlik1990}, that also simplifies to the Boubl\'ik-Mansoori-Carnahan-Starling-Leland equation of state\cite{Boublik1970,Mansoori1971,Carnahan1969} used in PCP-SAFT, can be used instead. The difference between the two models can be regarded as negligible compared to other model errors for the calculation of surface tensions.
The chain contribution $F^\mathrm{chain}[\bm{\rho}(\mathbf{r})]$, is a modified version of the functional by Tripathi and Chapman \cite{Tripathi2005,Tripathi2005a} for the PCP-SAFT equation of state. For the association contribution $F^\mathrm{assoc}[\bm{\rho}(\mathbf{r})]$, we use the model by Yu and Wu\cite{Yu2002b} and dispersive and polar interactions are combined in an attractive functional $F^\mathrm{att}[\bm{\rho}(\mathbf{r})]$, which uses a weighted density approach to account for the range of the interactions\cite{Sauer2017}. For the vector weight functions appearing in the FMT and association functionals, the expressions in Sec.~\ref{sec:DFT} have to be amended according to Appendix \ref{app:vector}. In previous works, the functional has already been applied to calculate the properties of adsorbed\cite{Sauer2018} and free droplets\cite{Rehner2018} as well as adsorption isotherms of pure components and mixtures\cite{Sauer2019}. With the exception of water, all components are described using parameters that have previously been published~\cite{Gross2001,Gross2005,Gross2006,Klink2014}.

\section{Results and discussion}
We first compare Helfrich coefficients obtained by use of different methodologies for pure components: the full non-local density functional theory as presented in this work, the predictive density gradient theory and standard density gradient theory (Sec.~\ref{sec:pure}). All theories reduce to the PCP-SAFT equation of state in bulk systems. While DFT and pDGT are predictive in nature, an influence parameter is required for DGT. There are various ways to obtain an appropriate influence parameter. However, since one of the objectives is to evaluate the influence of the Helmholtz energy functional on the Tolman lengths and rigidity constants, we set the DGT influence parameter to reproduce the surface tension of a planar interface predicted by the full DFT at each temperature.

Little is known about the Helfrich coefficients of mixtures. Because all derivations shown in Sec.~\ref{sec:theory} are valid for multicomponent systems, we use the DFT expressions together with the PCP-SAFT equation of state to examine the behavior of Helfrich coefficients in ideal and non-ideal mixtures in Sec.~\ref{sec:mix}.

All coefficients presented in the following are calculated using the surface of tension as dividing surface.

\subsection{Pure components \label{sec:pure}}
To confirm the validity of the calculated Helfrich coefficients and confirm the correctness of the implementation, we first compare the surface tension of droplets (positive curvature) and bubbles (negative curvatures) to results from the curvature expansion. For pDGT and DGT, the surface tensions are obtained by solving Eq. \eqref{eq:Noether} directly. For DFT we use the approach presented in a previous study\cite{Rehner2018}.

\begin{figure}
\includegraphics[width=\linewidth]{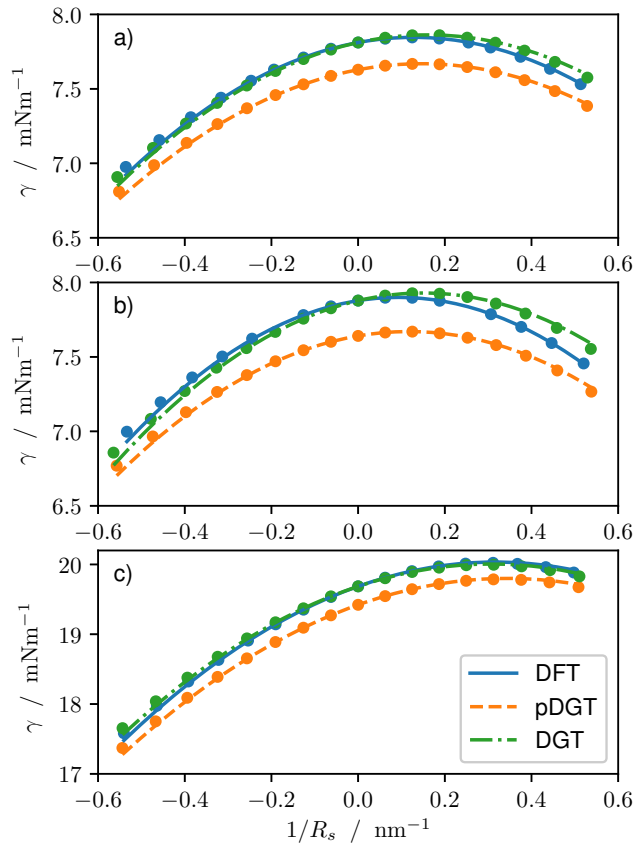}
\caption{Comparison of the surface tension of droplets ($R_s>0$) and bubbles ($R_s<0$) (symbols) with the Helfrich expansion (lines) for pDGT, DFT and DGT. The influence parameter in DGT is found by fixing the value of the planar surface tension to the corresponding result from DFT. a) methane at $T=140\,\text{K}$ b) n-hexane at $T=400\,\text{K}$, c) water at $T=550\,\text{K}$}
\label{fig:gamma_c}
\end{figure}

In Fig.~\ref{fig:gamma_c}, results from this comparison are shown for three different components and temperatures. In all cases, the surface tension of the droplets and bubbles is well approximated by the Helfrich expansion in the whole range of curvatures. In general, the different models also yield similar results, with the pDGT predicting slightly lower values for the surface tension for all curvatures. The planar surface tension from DGT is by construction equal to DFT. By increasing the curvature, the results start to differ with the effect being especially pronounced for n-hexane, the most elongated component considered.

\begin{figure}
\includegraphics[width=\linewidth]{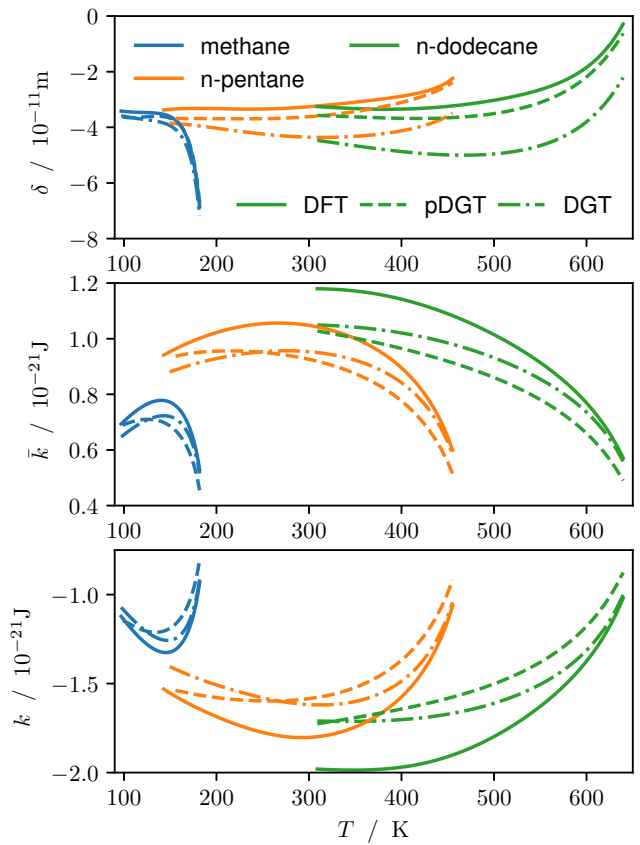}
\caption{Tolman length and rigidities of n-alkanes. Comparison between DFT, pDGT and DGT predictions.}
\label{fig:pure_alkanes}
\end{figure}

To obtain further insight about the chain length dependence of the Helfrich coefficients, we calculated them for n-alkanes of different lengths. Figure~\ref{fig:pure_alkanes} presents the results for methane, n-pentane and n-dodecane. Two observations made in previous work\cite{Rehner2018} can be confirmed here. The Tolman length of alkanes is over a wide temperature range very close to $-0.1$ times the segment diameter. In vicinity of the critical temperature however, the Tolman length deviates from this value. For small alkanes, the Tolman length decreases, whereas for longer alkanes the Tolman length increases. For methane, the different theories give similar results for the Tolman length. This conformity deteriorates for longer chains, with the magnitude of the DGT results being up to $50\,\%$ larger than the DFT results for n-dodecane.

A similar trend can be observed for the rigidity constants. The qualitative behavior is similar for all of the theories, but for longer chain lengths the difference between them increases. While the Tolman lengths from pDGT are close to the DFT results, both gradient based methods display comparable deviations from DFT for the rigidities, being up to $15\,\%$ for n-dodecane. Because bulk properties are described by the same equation of state for all considered theories, it is likely a difference in the description of structural properties that leads to the difference in predicted Helfrich coefficients. However, the way the different PCP-SAFT contributions affect the surface tension and the Helfrich coefficients is convoluted and not a simple linear combination. Hence, the role of the chain contribution in the different theories is not easily isolated. A more thorough investigation into the structure of interfaces of chain molecules, e.g. by molecular simulation, is advised to gain further insight about structural anisotropies at the interface.

% \begin{figure}
% \includegraphics[width=\linewidth]{Pure_polar}
% \caption{Tolman length and spherical rigidity of the polar components CO$_2$ and acetone. Comparison between DFT, pDGT and DGT predictions.}
% \label{fig:pure_polar}
% \end{figure}

\begin{figure}
\includegraphics[width=\linewidth]{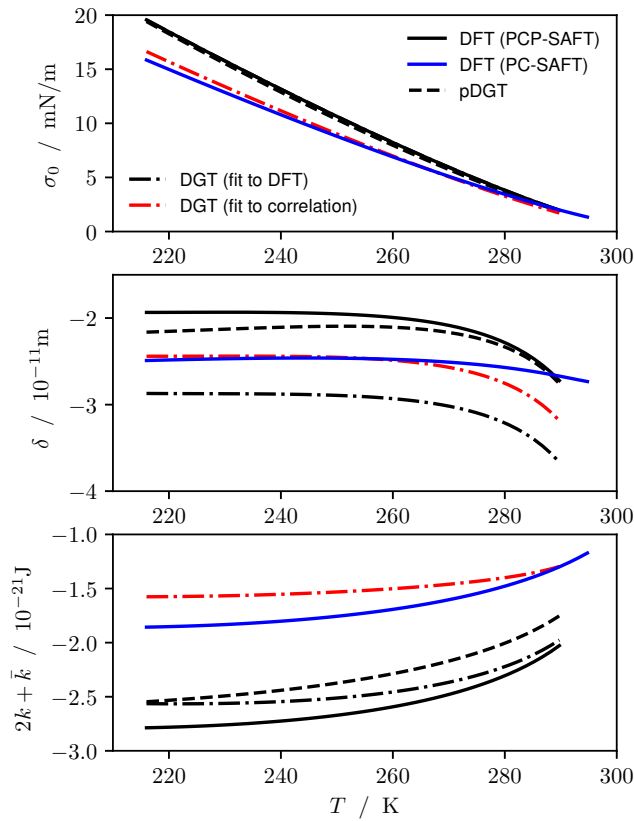}
\caption{Tolman length and spherical rigidity of CO$_2$. Comparison between DFT, pDGT and DGT predictions. Except for the blue line all results are obtained using the PCP-SAFT equation of state. For DGT, the results are obtained by fitting to the surface tension from DFT (black) and to an empirical correlation~\cite{Mulero2012} (red).}
\label{fig:pure_co2}
\end{figure}

To expand the study to polar components, the Helfrich coefficients of CO$_2$ are presented in Fig.~\ref{fig:pure_co2}. For homogeneous nucleation, the primary application of this framework, only spherical droplets are relevant. We find that the behavior of the two rigidities is very similar for all studied components. Therefore, from here on we only show the spherical rigidity $k_s=2k+\bar{k}$, which appears as the second order coefficient in a curvature expansion of the surface tension for a spherical geometry. The quantitative behavior of the different theories are similar for CO$_2$ and the alkanes. The predictions of the Tolman length from pDGT lie slightly below the DFT results with the difference decreasing with temperature. The DGT results on the other hand are significantly lower. For the rigidities, both pDGT and DGT predict larger values than DFT. We find that these are general trends for non-associating fluids, where results for other substances such as nitrogen and argon are included in the supplementary material.

Fig.~\ref{fig:pure_co2}-top shows that PC-SAFT predicts the surface tension of CO$_2$ to a reasonable accuracy, since DFT with PC-SAFT (blue solid line) agrees well with DGT for which the influence parameter was fitted to an empirical correlation~\cite{Mulero2012} for the surface tension (red dash-dot line). The surface tension from PCP-SAFT, however, lies above the experimental values. The values for the Tolman length and rigidities reflect this, where the Tolman length and rigidities from PC-SAFT and PCP-SAFT are significantly different. The prediction of the surface tension with the PC-SAFT equation of state is better than with PCP-SAFT, despite the latter describing the phase equilibrium and the critical point of CO$_2$ more accurately~\cite{Gross2006}. The same trend can be seen for DGT, where there is a large difference between the Helfrich coefficients when the influence parameter has been fitted to DFT values (black dash-dot line) and an empirical correlation (red dash-dot line). This effect is especially pronounced for the rigidity, which decreases about $40\,\%$ in magnitude by fitting to the empirical surface tension rather than to DFT. Hence, an important basis for reliable estimates of the curvature dependence of the surface tension is accurate prediction of the planar surface tension.

\begin{figure}
\includegraphics[width=\linewidth]{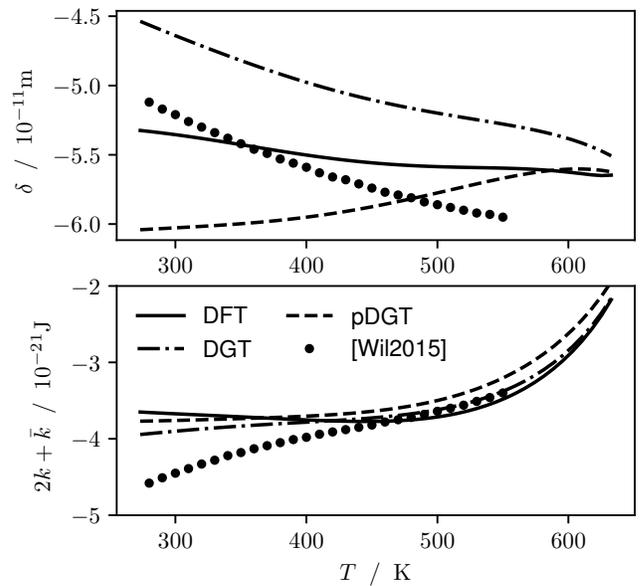}
\caption{
    Tolman length and spherical rigidity of water. Comparison between DFT, pDGT and DGT predictions for the PCP-SAFT equation of state, as well as DGT results using CPA from previous work\cite{Wilhelmsen2015a}.
    }
\end{figure}

We next discuss the Helfrich coefficients for water, as this has been a popular example in the literature~\cite{Wilhelmsen2015a,Menzl2016,Joswiak2016,Kim2018,Leong2018}. Since the numerous PCP-SAFT water parameter sets that have been published predict vastly different planar surface tensions\cite{Mairhofer2018}, new parameters have been obtained that use quasi experimental surface tension data\cite{Lemmon} as additional input in the estimation. These parameters are for the 2B association scheme\cite{Huang1990} and include a fitted dipole moment. They are shown in Tab.~\ref{tab:water_params}.

\begin{table}
\begin{tabular}{p{.23\columnwidth}p{.23\columnwidth}p{.23\columnwidth}p{.23\columnwidth}}
\hline
$m$ & .99214 & $\mu$ & 1.6152\,D\\
$\sigma$ & 3.0177\,\AA & $\kappa^{A_iB_i}$ & .091799 \\
$\varepsilon/k_B$ & 166.66\,K & $\varepsilon^{A_iB_i}/k_B$ & 2685.1\,K \\
\hline
\end{tabular}
\caption{PCP-SAFT parameters for the 2B water model used in this work.}
\label{tab:water_params}
\end{table}

For the Tolman length, we find the same behavior for pDGT and DFT as for non-associating fluids. The Tolman length obtained from DGT however, is larger than the DFT result. The spherical rigidity shows a remarkable resemblance for the three different approaches. We further compare the spherical rigidity to previous results\cite{Wilhelmsen2015a} that were calculated using DGT combined with the cubic plus association (CPA)~\cite{Kontogeorgis1996} equation of state. The Tolman length has a comparable magnitude and the temperature dependence is the same for DGT with PCP-SAFT. For the rigidity at higher temperatures, we again observe good agreement. For lower temperatures, the results deviate by up to $25\,\%$. Since the influence parameters do not differ significantly between the different approaches, this deviation can be attributed to the difference in equation of state. Hence, the equation of state has an important role in the prediction of the Helfrich coefficients. 

In conclusion, we find that the different descriptions of the considered Helmholtz energy functionals give relatively similar results. However, for strongly polar or elongated molecules, deviations between DFT and DGT should be expected, in particular for the Tolman length. Prerequisites for accurate prediction of the Helfrich coefficients are: a bulk equation of state that is able to describe the phase equilibrium well and a Helmholtz energy functional that is able to reproduce the planar surface tension accurately. As shown in the supplementary material, for alcohols, that are frequently used in nucleation experiments, the surface tension predictions using DFT and PCP-SAFT deviate significantly from experimental data. Therefore, further work has to be done to improve the parametrization of these components, before the influence of the curvature corrections on nucleation rates can be studied rigorously.

\subsection{Mixtures \label{sec:mix}}
In a previous work\cite{Aasen2018}, it was shown that the values of the Helfrich coefficients for mixtures are significantly influenced by the choice of path through the metastable region. We emphasize that already for pure components, a deliberate choice has been made by choosing the isothermal path. An isentropic path is another possible choice.

The value of the surface tension of a droplet is only a function of the thermodynamic state and the choice of dividing surface, and does not depend on the path. A different path, however, leads to a different quality of the prediction using the Helfrich expansion and a different composition dependence of the coefficients. Following the recommendations in a previous study by Aasen et al.\cite{Aasen2018}, we choose a straight path through the metastable region that keeps the liquid composition constant to first order.

\begin{figure}
\includegraphics[width=\linewidth]{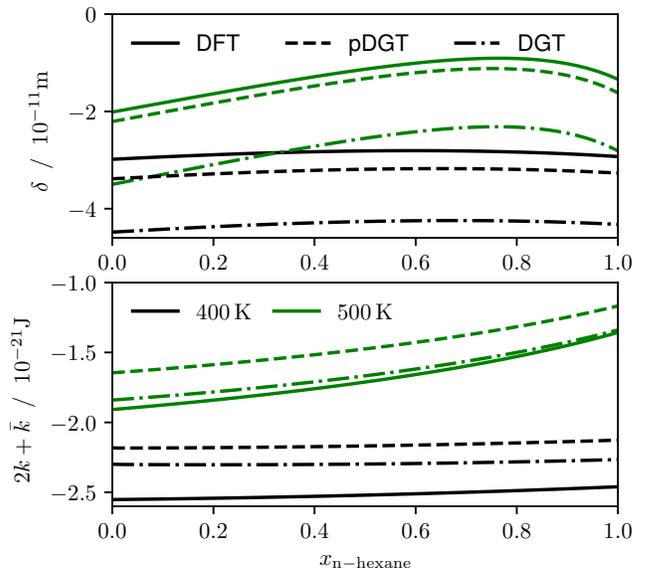}
\caption{Composition dependence of the Tolman length and spherical rigidity for the binary mixture of n-hexane and n-heptane at different temperatures. Comparison between DFT, pDGT and DGT results.}
\label{fig:mix_hex_hep}
\end{figure}

We first study the behavior of a close to ideal mixture. To that end, we examine the n-hexane/n-heptane mixture at different temperatures with the binary interaction parameter $k_{ij}$ equal to 0. In Fig.~\ref{fig:mix_hex_hep}, the Tolman length and the spherical rigidity are displayed as functions of the liquid mole fraction of n-hexane in the system. At lower temperatures, the pure component values of both coefficients are similar and there is almost no composition dependence. For temperatures close to the critical point, however, the Tolman length displays a non-linear dependence on the composition. The spherical rigidity is higher for n-hexane than for n-heptane closer to the critical point, but the composition dependence is still close to linear. Comparing the different theories, an almost constant deifference in predicted values can be observed over the whole composition range for both temperatures. Therefore, if a good agreement is obtained for the pure components, it can be expected that DGT using the geometric combining rule will also predict similar values as DFT and pDGT for this mixture.

\begin{figure}
\includegraphics[width=\linewidth]{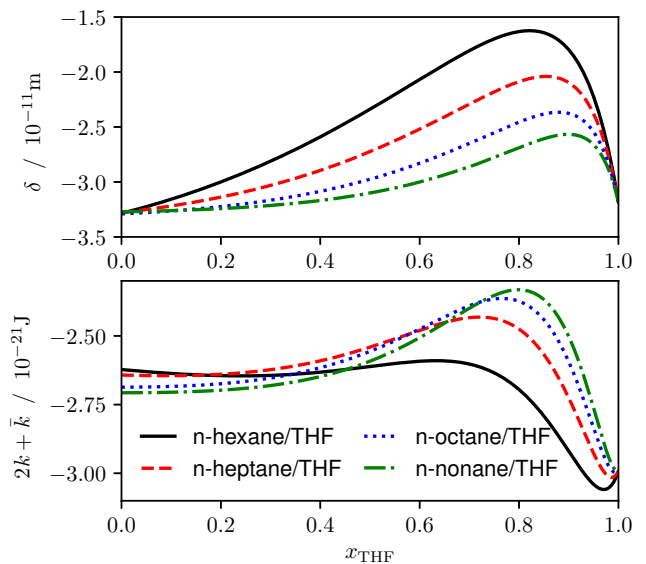}
\caption{DFT result for the composition dependence of the Tolman length and spherical rigidity for the binary mixture of various n-alkanes with the polar component tetrahydrofuran (THF) at $298.15\,\text{K}$.}
\label{fig:mix_thf}
\end{figure}

We extend the study to the more non ideal binary mixtures of n-alkanes with the polar solvent tetrahydrofuran (THF). Parameters for this system, including the binary interaction parameter $k_{ij}$ were obtained by Klink and Gross\cite{Klink2014} and the DFT results using them were shown to concur well with experimental data\cite{Sauer2017}. In Fig.~\ref{fig:mix_thf}, the Tolman length and spherical rigidity are shown for the binary mixture of THF with n-hexane, n-heptane, n-octane and n-nonane at $T=298.15\,\mathrm{K}$. Although all of the pure components have almost the same Tolman length at this temperature, the Tolman lengths of the mixtures are significantly different, with a peak near $x_{\text{THF}}=0.8$. This effect is most pronounced at higher concentrations of THF and for smaller alkanes, with the Tolman length of the n-hexane/THF mixture being up to $50\,\%$ higher than both pure component values. Contrary to that, the spherical rigidity of the same system is constant in a large composition range until the value drops towards the pure component value of THF. For longer alkanes, a peak in the spherical rigidity is observed, similar to the Tolman length. A comparison to the gradient theories can be seen in the supplementary material. Also for the non-ideal mixtures, DGT predicts a similar composition dependence as DFT for the Helfrich coefficients, with the main difference being a nearly constant difference in predicted values, which is determined by the deviation between the pure component values.

\section{Conclusion\label{sec:conclusion}}
The curvature dependence of the surface tension can be described by the Helfrich expansion, where the first and second order expansion coefficients are called the Tolman length and the rigidities. They are also referred to as the Helfrich coefficients.

In this work, we have derived general expressions that can be used for calculating Helfrich coefficients from any non-local Helmholtz energy functional based on weighted densities. The curvature expansion can be used to calculate the surface energy of arbitrarily shaped interfaces for pure component and mixtures.

We used the framework to compare predictions from non-local density functional theory (DFT) with results from density gradient theory (DGT) and predictive density gradient theory (pDGT). Good agreement between the different theories was observed for small, approximately spherical molecules. An increase in chain length led to larger differences in the predictions. We found that the values of the Helfrich coefficients are sensitive to the choice of influence parameter in DGT and to the prediction of the surface tension in DFT and pDGT. We showed that if a model is used that gives a good description of the phase equilibrium (including liquid densities) \textit{and} the surface tension, good agreement can be expected between the different functional theories. However, the Helfrich coefficients also displayed a dependence on the equation of state.

For non ideal mixtures, the composition dependence of the Helfrich coefficients was found to be nonlinear. All three functionals studied gave very similar composition dependencies for the Helfrich coefficients, where the difference comes mainly from different predictions of the pure component values.

Further work is needed to describe the Helfrich coefficients of alcohols, since PC-SAFT and PCP-SAFT are currently unable to accurately predict their surface tensions. Because alcohols are frequently used in nucleation experiments, their Helfrich coefficients are of much interest.

\section*{supplementary material}
See supplementary material for additional results for Helfrich coefficients, the derivation of the curvature expansion of convolution integrals and the simplification of the second order coefficient for DFT.

\begin{acknowledgments}
This work was partly supported by the Research Council of Norway through its Centres of Excellence funding scheme, project number 262644. P. Rehner acknowledges financial support from the German Research Foundation (DFG) through the collaborative research center \textit{Droplet Dynamics Under Extreme Ambient Conditions} (SFB-TRR 75) and from the German Academic Exchange Service (DAAD). We thank Joachim Gro\ss\ and Dick Bedeaux for helpful comments and discussions.
\end{acknowledgments}

\appendix
\section{Curvature expansion of convolution integrals\label{app:convolutions}}
In non-local DFT using weighted densities, the density profile and the partial derivatives of the Helmholtz energy density are convolved with a weight function $\omega$. To reduce the amount of different symbols, we use the same symbol for the different representations of $\omega$ and use the independent variable as an indicator for which representation to use. The different representations are the weight function in real space $\omega(\mathbf{r})=\omega(r)$, the projection on the $z$-axis
\begin{equation}
\omega(z)=2\pi\int\limits_{|z|}^\infty\omega(r)r\diff r
\end{equation}
and the Fourier transform
\begin{equation}
\omega(k)=\int\limits_{-\infty}^\infty\omega(z)e^{-2\pi ikz}\diff z=\int\limits_0^\infty\omega(r)\frac{2r}{k}\sin(2\pi kr)\diff r.
\label{eq:FourierOmega}
\end{equation}
The convolution of a spherically symmetric function $f(r)$ and a scalar weight function $\omega(r)$ can be expressed as\cite{Roth2010}
\begin{align}
f\stackrel{3D}{\otimes}\omega&=\frac{1}{r}\int f(r-z')(r-z')\omega(z')\diff z'\nonumber\\
&=f\otimes\omega-\frac{1}{r}f\otimes\left(z\omega\right)\nonumber\\
&=f\otimes\omega-f\otimes\left(z\omega\right)\frac{1}{R}+z\left(f\otimes\left(z\omega\right)\right)\frac{1}{R^2}+\ldots
\end{align}
The convolution of a cylindrically symmetric function with a scalar weight function is more intricate. The projection-slice theorem of the Fourier transform states, that the 3D Fourier transform can be replaced by a projection on one of the axes followed by the one-dimensional Fourier transform along the given axis. In a cylindrical geometry, the projection is known as the Abel transform. To our knowledge, no concise expression is available like in the spherical case. However, the curvature coefficients can still be derived by performing the curvature expansion on the general convolution integral itself. The expression we obtain is
\begin{multline}
f\stackrel{3D}{\otimes}\omega=f\otimes\omega-\frac{1}{2}f\otimes\left(z\omega\right)\frac{1}{R}\\
+\left(\frac{1}{2}z\left(f\otimes\left(z\omega\right)\right)-\frac{1}{8}f\otimes\tilde{\omega}\right)\frac{1}{R^2}+\ldots
\end{multline}
The full derivation is shown in the supplementary material.
% \begin{widetext}
% \begin{align}
% f\stackrel{3D}{\otimes}\omega&=\mathcal{A}^{-1}\mathcal{F}^{-1}\left(\mathcal{F}\mathcal{A}(f)\omega(k)\right)=-\frac{2}{\pi}\int\int\limits_{r-y'}^\infty\int\limits_r^{r'+y'}\frac{(y-y')}{\sqrt{(r'^2-(y-y')^2)(y^2-r^2)}}\diff yf'(r')\diff r'\omega(y')\diff y'\nonumber\\
% &=-\int\int\limits_{z-y'}^\infty\left(1-\frac{1}{2}y'\frac{1}{R}+\frac{1}{8}y'(3z+z')\frac{1}{R^2}+\ldots\right)f'(z')\diff z'\omega(y')\diff y'\nonumber\\
% &=f\otimes\omega-\frac{1}{2}f\otimes\left(z\omega\right)\frac{1}{R}+\left(\frac{1}{2}z\left(f\otimes\left(z\omega\right)\right)-\frac{1}{8}f\otimes\tilde{\omega}\right)\frac{1}{R^2}+\ldots
% \end{align}
% \end{widetext}
The weight function, $\tilde{\omega}$ appearing in the last convolution is
\begin{equation}
\tilde{\omega}=z^2\omega-\int\limits_z^\infty\omega(z')z'\diff z'=z^2\omega-\left(z\omega\right)\otimes\Theta(-z)
\end{equation}
with the Heaviside step function $\Theta(z)$. The two geometries can be combined in a general expression involving the geometry factor $g$, as
\begin{multline}
f\stackrel{3D}{\otimes}\omega=f\otimes\omega-\frac{g}{2}f\otimes\left(z\omega\right)\frac{1}{R}\\
+\left(\frac{g}{2}z\left(f\otimes\left(z\omega\right)\right)+\frac{g(g-2)}{8}f\otimes\tilde{\omega}\right)\frac{1}{R^2}+\ldots
\end{multline}

\section{Convolutions in Fourier space}
Aside from the convergence speed of the solver, the computation time of DFT is limited by the evaluation of the numerous convolution integrals. The calculation can be sped up using the convolution theorem of the Fourier transform. It states that the Fourier transform of a convolution is equal to the product of the Fourier transform of the functions that are being convolved. The Fourier transform of the density profiles and the partial derivatives can be calculated in $\mathcal{O}(N\log N)$ using the fast Fourier transform. The Fourier transform of the weight functions can be obtained analytically from Eq. \eqref{eq:FourierOmega}. The other weight functions needed to calculate the Helfrich coefficients can be obtained from the derivatives of the weight functions in Fourier space, as
\begin{align}
\mathcal{F}(z\omega)&=\frac{i}{2\pi}\omega'(k)~~~~~~\text{and}\\
\mathcal{F}(\tilde{\omega})&=\mathcal{F}\left(z^2\omega\right)-\mathcal{F}\left(z\omega\right)\mathcal{F}(\Theta(-z))\nonumber\\
&=-\frac{1}{4\pi^2}\omega''(k)-\frac{i}{2\pi}\omega'(k)\left(\frac{1}{2}\delta(k)+\frac{i}{2\pi k}\right)\nonumber\\
&=\frac{1}{4\pi^2}\left(\frac{\omega'(k)}{k}-\omega''(k)\right).
\end{align}
We focus on spherically symmetric weight functions, $\omega(r)$, $\omega(z)$ and $\omega(k)$ that are all even functions by construction. Therefore, $\omega'(k=0)=0$ and the term involving the dirac distribution $\delta(k)$ cancels. Further, using L'H\^opital's rule we find that $\mathcal{F}(\tilde{\omega})(k=0)=0$.

\section{Hyper-dual numbers}
With the only exception being the derivative $\bm{c}_{\bm{\rho} 0}$, all properties in the framework we discuss are related to at most second order partial derivatives of the Helmholtz energy density. Hyper-dual numbers\cite{Fike2011} can be used to calculate the exact second partial derivatives and thus all related properties. The approach has recently been used in the context of equations of state\cite{Diewald2018}. Here, we propose its use to calculate the first and second partial derivatives of the non-local Helmholtz energy density in DFT and to calculate the different weight functions in Fourier space needed to calculate all convolution integrals for the curvature expansion. Therefore, the only properties that need to be implemented are the Helmholtz energy functional and the weight functions. All other properties, including derivatives of the underlying equation of state and the weight constants in pDGT are available through the hyper-dual numbers, making it simpler and less error-prone to include new functionals. This improvement in usability comes with increased computation time, since every operator and intrinsic function has to be evaluated for hyper-dual numbers. In particular for functions of many variables, there is significant redundancy when calculating derivatives. Therefore, in cases with many variables and simple derivatives like the FMT and chain functionals, it is advisable to override the hyper-dual differentiation with analytic derivatives to speed up the computation.

\section{Vector weighted densities\label{app:vector}}
Some FMT\cite{Rosenfeld1989,Roth2002,Yu2002} and association functionals\cite{Yu2002b} use vector weighted densities. To include those in the framework presented in this work, the expressions have to be amended accordingly.
%If the vector weighted densities are as they appear in some FMT\cite{Rosenfeld1989,Roth2002,Yu2002} and association functionals\cite{Yu2002b}, the expressions in this work have to be amended accordingly. 
As we are still only considering spherically symmetric weight functions, we can write vector weight functions as $\vec{\omega}(\mathbf{r})=\omega_r(r)\vec{e}_r$ with the radial unit vector $\vec{e}_r$. The projection on the $z$-axis then becomes
\begin{equation}
\vec{\omega}(z)=\omega_z(z)\vec{e}_z=2\pi z\vec{e}_z\int\limits_{|z|}^\infty\omega_r(r)\diff r
\end{equation}
and the representation in Fourier space is
\begin{align}
\vec{\omega}(\mathbf{k})&=\omega_k(k)\vec{e}_k=\vec{e}_k\int\limits_{-\infty}^\infty\omega_z(z)e^{-2\pi ikz}\diff z\nonumber\\
&=\vec{e}_k\int\limits_0^\infty\omega_r(r)\frac{i}{\pi k^2}\left(2\pi kr\cos(2\pi kr)-\sin(2\pi kr)\right)\diff r.
\end{align}
The convolution integrals involving vector weight functions are different from scalar weight functions. A detailed derivation of the handling of these convolutions is given in the supporting material. To include vector weighted densities in the framework presented in this work, the convolution integrals in Sec. \ref{sec:DFT} have to be changed according to Tab.~\ref{tab:vector_weight_functions} for every vector weight function. The newly introduced weight function $\hat{\bm{\omega}}_{\alpha z}$ is defined as
\begin{equation}
\hat{\bm{\omega}}_{\alpha z}=\int\limits_z^\infty\bm{\omega}_{\alpha z}(z')\diff z'
\end{equation}
and all combinations of weight functions are again easily obtained in Fourier space as
\begin{align}
\mathcal{F}(z\omega_z\pm\hat{\omega}_z)&=\frac{i}{2\pi}\left(\omega_k'(k)\pm\frac{\omega_k(k)}{k}\right)\\
\mathcal{F}(z^2\omega_z\pm z\hat{\omega}_z)&=\frac{-1}{4\pi^2}\left(\omega_k''(k)\pm\left(\frac{\omega_k'(k)}{k}-\frac{\omega_k(k)}{k^2}\right)\right).
\end{align}

% Since the vector weight functions are odd functions by construction instead of even ones like the scalar weight functions, a sign change has to be taken into account in the Euler Lagrange equation
% \begin{equation}
% \bm{\mu}=\sum_\alpha^\mathrm{scal}f_{\alpha 0}\otimes\bm{\omega}_\alpha-\sum_\alpha^\mathrm{vec}\vec{f}_{\alpha 0}\otimes\vec{\bm{\omega}}_{\alpha}.
% \end{equation}
% The convolution $\bm{\rho}_0\otimes\left(z\bm{\omega}_\alpha\right)$ has to be replaced by
% \begin{align}
% \bm{\rho}_0\otimes\left(z\bm{\omega}_{\alpha z}-\hat{\bm{\omega}}_{\alpha z}\right)&&\text{with}&&\hat{\bm{\omega}}_{\alpha z}=\int\limits_z^\infty\bm{\omega}_{\alpha z}\diff z
% \end{align}
% whereas the convolution $f_{\alpha 0}\otimes\left(z\bm{\omega}_\alpha\right)$ is replaced by
% \begin{equation}
% -f_{\alpha 0}\otimes\left(z\bm{\omega}_{\alpha z}+\hat{\bm{\omega}}_{\alpha z}\right)
% \end{equation}
% The convolution $\bm{\rho}_0\otimes\tilde{\bm{\omega}}_\alpha$ in the expression for $k$ is rewritten as
% \begin{align}
% TBD
% \end{align}
% and the same convolution in the expression for $\bar{k}$ is written as
% \begin{align}
% TBD.
% \end{align}
% A detailed derivation of these expressions is given in the supplementary material. Again, all different forms of the weight functions can be conveniently calculated in Fourier space.

\begin{table}
\begin{tabular}{lll}
scalar expression & vector expression & equations \\
\hline
$\bm{\rho}_{0|1}\otimes\bm{\omega}_\alpha$ & $\bm{\rho}_{0|1}\otimes\bm{\omega}_{\alpha z}$ & \eqref{eq:n_alpha_0}, \eqref{eq:n_alpha_1}\\
$f_{\alpha 0|1}\otimes\bm{\omega}_\alpha$ & $-f_{\alpha 0|1}\otimes\bm{\omega}_{\alpha z}$ & \eqref{eq:EulerLagrange0}, \eqref{eq:EulerLagrange1} \\
$\bm{\rho}_0\otimes\left(z\bm{\omega}_\alpha\right)$ & $\bm{\rho}_0\otimes\left(z\bm{\omega}_{\alpha z}-\hat{\bm{\omega}}_{\alpha z}\right)$ & \eqref{eq:Tolman_DFT} - \eqref{eq:kbar_DFT}, \eqref{eq:n_alpha_1} \\
$f_{\alpha 0}\otimes\left(z\bm{\omega}_\alpha\right)$ & $-f_{\alpha 0}\otimes\left(z\bm{\omega}_{\alpha z}+\hat{\bm{\omega}}_{\alpha z}\right)$ & \eqref{eq:k_DFT}, \eqref{eq:EulerLagrange1} \\
$\bm{\rho}_0\otimes\tilde{\bm{\omega}}_\alpha$ & $\bm{\rho}_0\otimes\left(z^2\bm{\omega}_{\alpha z}+z\hat{\bm{\omega}}_{\alpha z}\right)$ & \eqref{eq:k_DFT} \\
$\bm{\rho}_0\otimes\tilde{\bm{\omega}}_\alpha$ & $\bm{\rho}_0\otimes\left(z^2\bm{\omega}_{\alpha z}-z\hat{\bm{\omega}}_{\alpha z}\right)$ & \eqref{eq:kbar_DFT} \\
\end{tabular}
\caption{Replacement for convolution integrals for vector weighted densities.}
\label{tab:vector_weight_functions}
\end{table}

\bibliography{../../../../Literatur/Literatur.bib}

\begin{thebibliography}{66}%
\makeatletter
\providecommand \@ifxundefined [1]{%
 \@ifx{#1\undefined}
}%
\providecommand \@ifnum [1]{%
 \ifnum #1\expandafter \@firstoftwo
 \else \expandafter \@secondoftwo
 \fi
}%
\providecommand \@ifx [1]{%
 \ifx #1\expandafter \@firstoftwo
 \else \expandafter \@secondoftwo
 \fi
}%
\providecommand \natexlab [1]{#1}%
\providecommand \enquote  [1]{``#1''}%
\providecommand \bibnamefont  [1]{#1}%
\providecommand \bibfnamefont [1]{#1}%
\providecommand \citenamefont [1]{#1}%
\providecommand \href@noop [0]{\@secondoftwo}%
\providecommand \href [0]{\begingroup \@sanitize@url \@href}%
\providecommand \@href[1]{\@@startlink{#1}\@@href}%
\providecommand \@@href[1]{\endgroup#1\@@endlink}%
\providecommand \@sanitize@url [0]{\catcode `\\12\catcode `\$12\catcode
  `\&12\catcode `\#12\catcode `\^12\catcode `\_12\catcode `\%12\relax}%
\providecommand \@@startlink[1]{}%
\providecommand \@@endlink[0]{}%
\providecommand \url  [0]{\begingroup\@sanitize@url \@url }%
\providecommand \@url [1]{\endgroup\@href {#1}{\urlprefix }}%
\providecommand \urlprefix  [0]{URL }%
\providecommand \Eprint [0]{\href }%
\providecommand \doibase [0]{http://dx.doi.org/}%
\providecommand \selectlanguage [0]{\@gobble}%
\providecommand \bibinfo  [0]{\@secondoftwo}%
\providecommand \bibfield  [0]{\@secondoftwo}%
\providecommand \translation [1]{[#1]}%
\providecommand \BibitemOpen [0]{}%
\providecommand \bibitemStop [0]{}%
\providecommand \bibitemNoStop [0]{.\EOS\space}%
\providecommand \EOS [0]{\spacefactor3000\relax}%
\providecommand \BibitemShut  [1]{\csname bibitem#1\endcsname}%
\let\auto@bib@innerbib\@empty
%</preamble>
\bibitem [{\citenamefont {Iland}\ \emph {et~al.}(2007)\citenamefont {Iland},
  \citenamefont {W\"{o}lk}, \citenamefont {Strey},\ and\ \citenamefont
  {Kashchiev}}]{Iland2007}%
  \BibitemOpen
  \bibfield  {author} {\bibinfo {author} {\bibfnamefont {K.}~\bibnamefont
  {Iland}}, \bibinfo {author} {\bibfnamefont {J.}~\bibnamefont {W\"{o}lk}},
  \bibinfo {author} {\bibfnamefont {R.}~\bibnamefont {Strey}}, \ and\ \bibinfo
  {author} {\bibfnamefont {D.}~\bibnamefont {Kashchiev}},\ }\bibfield  {title}
  {\enquote {\bibinfo {title} {{Argon Nucleation in a Cryogenic Pulse
  Chamber}},}\ }\href@noop {} {\bibfield  {journal} {\bibinfo  {journal}
  {Journal of Chemical Physics}\ }\textbf {\bibinfo {volume} {127}},\ \bibinfo
  {pages} {154506} (\bibinfo {year} {2007})}\BibitemShut {NoStop}%
\bibitem [{\citenamefont {Vehkam{\"a}ki}(2006)}]{Vehkamaeki2006}%
  \BibitemOpen
  \bibfield  {author} {\bibinfo {author} {\bibfnamefont {H.}~\bibnamefont
  {Vehkam{\"a}ki}},\ }\href@noop {} {\emph {\bibinfo {title} {Classical
  nucleation theory in multicomponent systems}}}\ (\bibinfo  {publisher}
  {Springer Science \& Business Media},\ \bibinfo {year} {2006})\BibitemShut
  {NoStop}%
\bibitem [{\citenamefont {ten Wolde}\ and\ \citenamefont
  {Frenkel}(1998)}]{Wolde1998a}%
  \BibitemOpen
  \bibfield  {author} {\bibinfo {author} {\bibfnamefont {P.~R.}\ \bibnamefont
  {ten Wolde}}\ and\ \bibinfo {author} {\bibfnamefont {D.}~\bibnamefont
  {Frenkel}},\ }\bibfield  {title} {\enquote {\bibinfo {title} {{Computer
  simulation study of gas-liquid nucleation in a Lennard-Jones system}},}\
  }\href {\doibase 10.1063/1.477658} {\bibfield  {journal} {\bibinfo  {journal}
  {The Journal of Chemical Physics}\ }\textbf {\bibinfo {volume} {109}},\
  \bibinfo {pages} {9901} (\bibinfo {year} {1998})}\BibitemShut {NoStop}%
\bibitem [{\citenamefont {Wilhelmsen}, \citenamefont {Bedeaux},\ and\
  \citenamefont {Reguera}(2015{\natexlab{a}})}]{Wilhelmsen2015a}%
  \BibitemOpen
  \bibfield  {author} {\bibinfo {author} {\bibfnamefont {{\O}.}~\bibnamefont
  {Wilhelmsen}}, \bibinfo {author} {\bibfnamefont {D.}~\bibnamefont {Bedeaux}},
  \ and\ \bibinfo {author} {\bibfnamefont {D.}~\bibnamefont {Reguera}},\
  }\bibfield  {title} {\enquote {\bibinfo {title} {Communication: Tolman length
  and rigidity constants of water and their role in nucleation},}\ }\href
  {\doibase 10.1063/1.4919689} {\bibfield  {journal} {\bibinfo  {journal} {The
  Journal of Chemical Physics}\ }\textbf {\bibinfo {volume} {142}},\ \bibinfo
  {pages} {171103} (\bibinfo {year} {2015}{\natexlab{a}})}\BibitemShut
  {NoStop}%
\bibitem [{\citenamefont {Nguyen}\ \emph {et~al.}(2018)\citenamefont {Nguyen},
  \citenamefont {Schoemaker}, \citenamefont {Blokhuis},\ and\ \citenamefont
  {Schall}}]{Nguyen2018}%
  \BibitemOpen
  \bibfield  {author} {\bibinfo {author} {\bibfnamefont {V.~D.}\ \bibnamefont
  {Nguyen}}, \bibinfo {author} {\bibfnamefont {F.~C.}\ \bibnamefont
  {Schoemaker}}, \bibinfo {author} {\bibfnamefont {E.~M.}\ \bibnamefont
  {Blokhuis}}, \ and\ \bibinfo {author} {\bibfnamefont {P.}~\bibnamefont
  {Schall}},\ }\bibfield  {title} {\enquote {\bibinfo {title} {{Measurement of
  the Curvature-Dependent Surface Tension in Nucleating Colloidal Liquids}},}\
  }\href {\doibase 10.1103/PhysRevLett.121.246102} {\bibfield  {journal}
  {\bibinfo  {journal} {Physical Review Letters}\ }\textbf {\bibinfo {volume}
  {121}},\ \bibinfo {pages} {246102} (\bibinfo {year} {2018})}\BibitemShut
  {NoStop}%
\bibitem [{\citenamefont {Helfrich}(1973)}]{Helfrich1973}%
  \BibitemOpen
  \bibfield  {author} {\bibinfo {author} {\bibfnamefont {W.}~\bibnamefont
  {Helfrich}},\ }\bibfield  {title} {\enquote {\bibinfo {title} {Elastic
  properties of lipid bilayers: Theory and possible experiments},}\ }\href
  {\doibase 10.1515/znc-1973-11-1209} {\bibfield  {journal} {\bibinfo
  {journal} {Zeitschrift für Naturforschung C}\ }\textbf {\bibinfo {volume}
  {28}} (\bibinfo {year} {1973}),\ 10.1515/znc-1973-11-1209}\BibitemShut
  {NoStop}%
\bibitem [{\citenamefont {Kim}\ \emph {et~al.}(2018)\citenamefont {Kim},
  \citenamefont {Kim}, \citenamefont {Kim}, \citenamefont {An},\ and\
  \citenamefont {Jhe}}]{Kim2018}%
  \BibitemOpen
  \bibfield  {author} {\bibinfo {author} {\bibfnamefont {S.}~\bibnamefont
  {Kim}}, \bibinfo {author} {\bibfnamefont {D.}~\bibnamefont {Kim}}, \bibinfo
  {author} {\bibfnamefont {J.}~\bibnamefont {Kim}}, \bibinfo {author}
  {\bibfnamefont {S.}~\bibnamefont {An}}, \ and\ \bibinfo {author}
  {\bibfnamefont {W.}~\bibnamefont {Jhe}},\ }\bibfield  {title} {\enquote
  {\bibinfo {title} {Direct evidence for curvature-dependent surface tension in
  capillary condensation: Kelvin equation at molecular scale},}\ }\href
  {\doibase 10.1103/PhysRevX.8.041046} {\bibfield  {journal} {\bibinfo
  {journal} {Physical Review X}\ }\textbf {\bibinfo {volume} {8}} (\bibinfo
  {year} {2018}),\ 10.1103/PhysRevX.8.041046}\BibitemShut {NoStop}%
\bibitem [{\citenamefont {Tolman}(1949)}]{Tolman1949}%
  \BibitemOpen
  \bibfield  {author} {\bibinfo {author} {\bibfnamefont {R.~C.}\ \bibnamefont
  {Tolman}},\ }\bibfield  {title} {\enquote {\bibinfo {title} {The effect of
  droplet size on surface tension},}\ }\href {\doibase 10.1063/1.1747247}
  {\bibfield  {journal} {\bibinfo  {journal} {The Journal of Chemical Physics}\
  }\textbf {\bibinfo {volume} {17}},\ \bibinfo {pages} {333--337} (\bibinfo
  {year} {1949})}\BibitemShut {NoStop}%
\bibitem [{\citenamefont {Wilhelmsen}, \citenamefont {Bedeaux},\ and\
  \citenamefont {Reguera}(2015{\natexlab{b}})}]{Wilhelmsen2015}%
  \BibitemOpen
  \bibfield  {author} {\bibinfo {author} {\bibfnamefont {{\O}.}~\bibnamefont
  {Wilhelmsen}}, \bibinfo {author} {\bibfnamefont {D.}~\bibnamefont {Bedeaux}},
  \ and\ \bibinfo {author} {\bibfnamefont {D.}~\bibnamefont {Reguera}},\
  }\bibfield  {title} {\enquote {\bibinfo {title} {Tolman length and rigidity
  constants of the lennard-jones fluid},}\ }\href {\doibase 10.1063/1.4907588}
  {\bibfield  {journal} {\bibinfo  {journal} {The Journal of Chemical Physics}\
  }\textbf {\bibinfo {volume} {142}},\ \bibinfo {pages} {064706} (\bibinfo
  {year} {2015}{\natexlab{b}})}\BibitemShut {NoStop}%
\bibitem [{\citenamefont {Bruot}\ and\ \citenamefont
  {Caupin}(2016)}]{Bruot2016}%
  \BibitemOpen
  \bibfield  {author} {\bibinfo {author} {\bibfnamefont {N.}~\bibnamefont
  {Bruot}}\ and\ \bibinfo {author} {\bibfnamefont {F.}~\bibnamefont {Caupin}},\
  }\bibfield  {title} {\enquote {\bibinfo {title} {Curvature dependence of the
  liquid-vapor surface tension beyond the tolman approximation},}\ }\href
  {\doibase 10.1103/PhysRevLett.116.056102} {\bibfield  {journal} {\bibinfo
  {journal} {Physical Review Letters}\ }\textbf {\bibinfo {volume} {116}},\
  \bibinfo {pages} {056102} (\bibinfo {year} {2016})}\BibitemShut {NoStop}%
\bibitem [{\citenamefont {Aasen}, \citenamefont {Blokhuis},\ and\ \citenamefont
  {Wilhelmsen}(2018)}]{Aasen2018}%
  \BibitemOpen
  \bibfield  {author} {\bibinfo {author} {\bibfnamefont {A.}~\bibnamefont
  {Aasen}}, \bibinfo {author} {\bibfnamefont {E.~M.}\ \bibnamefont {Blokhuis}},
  \ and\ \bibinfo {author} {\bibfnamefont {Ã.}~\bibnamefont {Wilhelmsen}},\
  }\bibfield  {title} {\enquote {\bibinfo {title} {Tolman lengths and rigidity
  constants of multicomponent fluids: Fundamental theory and numerical
  examples},}\ }\href {\doibase 10.1063/1.5026747} {\bibfield  {journal}
  {\bibinfo  {journal} {The Journal of Chemical Physics}\ }\textbf {\bibinfo
  {volume} {148}},\ \bibinfo {pages} {204702} (\bibinfo {year}
  {2018})}\BibitemShut {NoStop}%
\bibitem [{\citenamefont {Rehner}\ and\ \citenamefont
  {Gross}(2018{\natexlab{a}})}]{Rehner2018}%
  \BibitemOpen
  \bibfield  {author} {\bibinfo {author} {\bibfnamefont {P.}~\bibnamefont
  {Rehner}}\ and\ \bibinfo {author} {\bibfnamefont {J.}~\bibnamefont {Gross}},\
  }\bibfield  {title} {\enquote {\bibinfo {title} {Surface tension of droplets
  and tolman lengths of real substances and mixtures from density functional
  theory},}\ }\href {\doibase 10.1063/1.5020421} {\bibfield  {journal}
  {\bibinfo  {journal} {The Journal of Chemical Physics}\ }\textbf {\bibinfo
  {volume} {148}},\ \bibinfo {pages} {164703} (\bibinfo {year}
  {2018}{\natexlab{a}})}\BibitemShut {NoStop}%
\bibitem [{\citenamefont {Blokhuis}\ and\ \citenamefont
  {Kuipers}(2006)}]{Blokhuis2006}%
  \BibitemOpen
  \bibfield  {author} {\bibinfo {author} {\bibfnamefont {E.~M.}\ \bibnamefont
  {Blokhuis}}\ and\ \bibinfo {author} {\bibfnamefont {J.}~\bibnamefont
  {Kuipers}},\ }\bibfield  {title} {\enquote {\bibinfo {title} {Thermodynamic
  expressions for the tolman length},}\ }\href {\doibase 10.1063/1.2167642}
  {\bibfield  {journal} {\bibinfo  {journal} {Journal of Chemical Physics}\
  }\textbf {\bibinfo {volume} {124}},\ \bibinfo {pages} {74701--74701}
  (\bibinfo {year} {2006})}\BibitemShut {NoStop}%
\bibitem [{\citenamefont {Li}\ and\ \citenamefont {Wu}(2007)}]{Li2007}%
  \BibitemOpen
  \bibfield  {author} {\bibinfo {author} {\bibfnamefont {Z.}~\bibnamefont
  {Li}}\ and\ \bibinfo {author} {\bibfnamefont {J.}~\bibnamefont {Wu}},\
  }\bibfield  {title} {\enquote {\bibinfo {title} {Toward a quantitative theory
  of ultrasmall liquid droplets and vapor liquid nucleation},}\ }\href
  {\doibase 10.1021/ie070578i} {\bibfield  {journal} {\bibinfo  {journal}
  {Industrial \& Engineering Chemistry Research}\ }\textbf {\bibinfo {volume}
  {47}},\ \bibinfo {pages} {4988--4995} (\bibinfo {year} {2007})}\BibitemShut
  {NoStop}%
\bibitem [{\citenamefont {Block}\ \emph {et~al.}(2010)\citenamefont {Block},
  \citenamefont {Das}, \citenamefont {Oettel}, \citenamefont {Virnau},\ and\
  \citenamefont {Binder}}]{Block2010}%
  \BibitemOpen
  \bibfield  {author} {\bibinfo {author} {\bibfnamefont {B.~J.}\ \bibnamefont
  {Block}}, \bibinfo {author} {\bibfnamefont {S.~K.}\ \bibnamefont {Das}},
  \bibinfo {author} {\bibfnamefont {M.}~\bibnamefont {Oettel}}, \bibinfo
  {author} {\bibfnamefont {P.}~\bibnamefont {Virnau}}, \ and\ \bibinfo {author}
  {\bibfnamefont {K.}~\bibnamefont {Binder}},\ }\bibfield  {title} {\enquote
  {\bibinfo {title} {Curvature dependence of surface free energy of liquid
  drops and bubbles: A simulation study},}\ }\href {\doibase 10.1063/1.3493464}
  {\bibfield  {journal} {\bibinfo  {journal} {The Journal of Chemical Physics}\
  }\textbf {\bibinfo {volume} {133}},\ \bibinfo {pages} {154702} (\bibinfo
  {year} {2010})}\BibitemShut {NoStop}%
\bibitem [{\citenamefont {Tr{\"o}ster}\ \emph {et~al.}(2012)\citenamefont
  {Tr{\"o}ster}, \citenamefont {Oettel}, \citenamefont {Block}, \citenamefont
  {Virnau},\ and\ \citenamefont {Binder}}]{Troester2012}%
  \BibitemOpen
  \bibfield  {author} {\bibinfo {author} {\bibfnamefont {A.}~\bibnamefont
  {Tr{\"o}ster}}, \bibinfo {author} {\bibfnamefont {M.}~\bibnamefont {Oettel}},
  \bibinfo {author} {\bibfnamefont {B.}~\bibnamefont {Block}}, \bibinfo
  {author} {\bibfnamefont {P.}~\bibnamefont {Virnau}}, \ and\ \bibinfo {author}
  {\bibfnamefont {K.}~\bibnamefont {Binder}},\ }\bibfield  {title} {\enquote
  {\bibinfo {title} {Numerical approaches to determine the interface tension of
  curved interfaces from free energy calculations},}\ }\href {\doibase
  10.1063/1.3685221} {\bibfield  {journal} {\bibinfo  {journal} {The Journal of
  Chemical Physics}\ }\textbf {\bibinfo {volume} {136}},\ \bibinfo {pages}
  {064709} (\bibinfo {year} {2012})}\BibitemShut {NoStop}%
\bibitem [{\citenamefont {van Giessen}\ and\ \citenamefont
  {Blokhuis}(2002)}]{Giessen2002}%
  \BibitemOpen
  \bibfield  {author} {\bibinfo {author} {\bibfnamefont {A.~E.}\ \bibnamefont
  {van Giessen}}\ and\ \bibinfo {author} {\bibfnamefont {E.~M.}\ \bibnamefont
  {Blokhuis}},\ }\bibfield  {title} {\enquote {\bibinfo {title} {Determination
  of curvature corrections to the surface tension of a liquid–vapor interface
  through molecular dynamics simulations},}\ }\href {\doibase
  10.1063/1.1423617} {\bibfield  {journal} {\bibinfo  {journal} {The Journal of
  Chemical Physics}\ }\textbf {\bibinfo {volume} {116}},\ \bibinfo {pages}
  {302--310} (\bibinfo {year} {2002})}\BibitemShut {NoStop}%
\bibitem [{\citenamefont {Lei}\ \emph {et~al.}(2005)\citenamefont {Lei},
  \citenamefont {Bykov}, \citenamefont {Yoo},\ and\ \citenamefont
  {Zeng}}]{Lei2005}%
  \BibitemOpen
  \bibfield  {author} {\bibinfo {author} {\bibfnamefont {Y.~A.}\ \bibnamefont
  {Lei}}, \bibinfo {author} {\bibfnamefont {T.}~\bibnamefont {Bykov}}, \bibinfo
  {author} {\bibfnamefont {S.}~\bibnamefont {Yoo}}, \ and\ \bibinfo {author}
  {\bibfnamefont {X.~C.}\ \bibnamefont {Zeng}},\ }\bibfield  {title} {\enquote
  {\bibinfo {title} {{The Tolman Length: Is it Positive or Negative?}}}\
  }\href@noop {} {\bibfield  {journal} {\bibinfo  {journal} {J. Am. Chem.
  Soc.}\ }\textbf {\bibinfo {volume} {127}},\ \bibinfo {pages} {15346}
  (\bibinfo {year} {2005})}\BibitemShut {NoStop}%
\bibitem [{\citenamefont {Horsch}\ \emph {et~al.}(2012)\citenamefont {Horsch},
  \citenamefont {Hasse}, \citenamefont {Shchekin}, \citenamefont {Agarwal},
  \citenamefont {Eckelsbach}, \citenamefont {Vrabec}, \citenamefont
  {M\"uller},\ and\ \citenamefont {Jackson}}]{Horsch2012}%
  \BibitemOpen
  \bibfield  {author} {\bibinfo {author} {\bibfnamefont {M.}~\bibnamefont
  {Horsch}}, \bibinfo {author} {\bibfnamefont {H.}~\bibnamefont {Hasse}},
  \bibinfo {author} {\bibfnamefont {A.~K.}\ \bibnamefont {Shchekin}}, \bibinfo
  {author} {\bibfnamefont {A.}~\bibnamefont {Agarwal}}, \bibinfo {author}
  {\bibfnamefont {S.}~\bibnamefont {Eckelsbach}}, \bibinfo {author}
  {\bibfnamefont {J.}~\bibnamefont {Vrabec}}, \bibinfo {author} {\bibfnamefont
  {E.~A.}\ \bibnamefont {M\"uller}}, \ and\ \bibinfo {author} {\bibfnamefont
  {G.}~\bibnamefont {Jackson}},\ }\bibfield  {title} {\enquote {\bibinfo
  {title} {Excess equimolar radius of liquid drops},}\ }\href {\doibase
  10.1103/PhysRevE.85.031605} {\bibfield  {journal} {\bibinfo  {journal}
  {Physical Review E}\ }\textbf {\bibinfo {volume} {85}},\ \bibinfo {pages}
  {031605} (\bibinfo {year} {2012})}\BibitemShut {NoStop}%
\bibitem [{\citenamefont {van Giessen}\ and\ \citenamefont
  {Blokhuis}(2009)}]{Giessen2009}%
  \BibitemOpen
  \bibfield  {author} {\bibinfo {author} {\bibfnamefont {A.~E.}\ \bibnamefont
  {van Giessen}}\ and\ \bibinfo {author} {\bibfnamefont {E.~M.}\ \bibnamefont
  {Blokhuis}},\ }\bibfield  {title} {\enquote {\bibinfo {title} {Direct
  determination of the tolman length from the bulk pressures of liquid drops
  via molecular dynamics simulations},}\ }\href {\doibase 10.1063/1.3253685}
  {\bibfield  {journal} {\bibinfo  {journal} {The Journal of Chemical Physics}\
  }\textbf {\bibinfo {volume} {131}},\ \bibinfo {eid} {164705} (\bibinfo {year}
  {2009})}\BibitemShut {NoStop}%
\bibitem [{\citenamefont {Sampayo}\ \emph {et~al.}(2010)\citenamefont
  {Sampayo}, \citenamefont {Malijevsk{\'y}}, \citenamefont {M{\"u}ller},
  \citenamefont {de~Miguel},\ and\ \citenamefont {Jackson}}]{Sampayo2010}%
  \BibitemOpen
  \bibfield  {author} {\bibinfo {author} {\bibfnamefont {J.~G.}\ \bibnamefont
  {Sampayo}}, \bibinfo {author} {\bibfnamefont {A.}~\bibnamefont
  {Malijevsk{\'y}}}, \bibinfo {author} {\bibfnamefont {E.~A.}\ \bibnamefont
  {M{\"u}ller}}, \bibinfo {author} {\bibfnamefont {E.}~\bibnamefont
  {de~Miguel}}, \ and\ \bibinfo {author} {\bibfnamefont {G.}~\bibnamefont
  {Jackson}},\ }\bibfield  {title} {\enquote {\bibinfo {title} {Communications:
  Evidence for the role of fluctuations in the thermodynamics of nanoscale
  drops and the implications in computations of the surface tension},}\ }\href
  {\doibase 10.1063/1.3376612} {\bibfield  {journal} {\bibinfo  {journal} {The
  Journal of Chemical Physics}\ }\textbf {\bibinfo {volume} {132}},\ \bibinfo
  {eid} {141101} (\bibinfo {year} {2010})}\BibitemShut {NoStop}%
\bibitem [{\citenamefont {Blokhuis}\ and\ \citenamefont {van
  Giessen}(2013)}]{Blokhuis2013}%
  \BibitemOpen
  \bibfield  {author} {\bibinfo {author} {\bibfnamefont {E.~M.}\ \bibnamefont
  {Blokhuis}}\ and\ \bibinfo {author} {\bibfnamefont {A.~E.}\ \bibnamefont {van
  Giessen}},\ }\bibfield  {title} {\enquote {\bibinfo {title} {Density
  functional theory of a curved liquid-vapour interface: evaluation of the
  rigidity constants},}\ }\href {\doibase 10.1088/0953-8984/25/22/225003}
  {\bibfield  {journal} {\bibinfo  {journal} {Journal of Physics: Condensed
  Matter}\ }\textbf {\bibinfo {volume} {25}},\ \bibinfo {pages} {225003}
  (\bibinfo {year} {2013})}\BibitemShut {NoStop}%
\bibitem [{\citenamefont {Menzl}\ \emph {et~al.}(2016)\citenamefont {Menzl},
  \citenamefont {Gonzalez}, \citenamefont {Geiger}, \citenamefont {Caupin},
  \citenamefont {Abascal}, \citenamefont {Valeriani},\ and\ \citenamefont
  {Dellago}}]{Menzl2016}%
  \BibitemOpen
  \bibfield  {author} {\bibinfo {author} {\bibfnamefont {G.}~\bibnamefont
  {Menzl}}, \bibinfo {author} {\bibfnamefont {M.~A.}\ \bibnamefont {Gonzalez}},
  \bibinfo {author} {\bibfnamefont {P.}~\bibnamefont {Geiger}}, \bibinfo
  {author} {\bibfnamefont {F.}~\bibnamefont {Caupin}}, \bibinfo {author}
  {\bibfnamefont {J.~L.~F.}\ \bibnamefont {Abascal}}, \bibinfo {author}
  {\bibfnamefont {C.}~\bibnamefont {Valeriani}}, \ and\ \bibinfo {author}
  {\bibfnamefont {C.}~\bibnamefont {Dellago}},\ }\bibfield  {title} {\enquote
  {\bibinfo {title} {Molecular mechanism for cavitation in water under
  tension},}\ }\href {\doibase 10.1073/pnas.1608421113} {\bibfield  {journal}
  {\bibinfo  {journal} {Proceedings of the National Academy of Sciences}\
  }\textbf {\bibinfo {volume} {113}},\ \bibinfo {pages} {13582--13587}
  (\bibinfo {year} {2016})}\BibitemShut {NoStop}%
\bibitem [{\citenamefont {Joswiak}\ \emph {et~al.}(2013)\citenamefont
  {Joswiak}, \citenamefont {Duff}, \citenamefont {Doherty},\ and\ \citenamefont
  {Peters}}]{Joswiak2013}%
  \BibitemOpen
  \bibfield  {author} {\bibinfo {author} {\bibfnamefont {M.~N.}\ \bibnamefont
  {Joswiak}}, \bibinfo {author} {\bibfnamefont {N.}~\bibnamefont {Duff}},
  \bibinfo {author} {\bibfnamefont {M.~F.}\ \bibnamefont {Doherty}}, \ and\
  \bibinfo {author} {\bibfnamefont {B.}~\bibnamefont {Peters}},\ }\bibfield
  {title} {\enquote {\bibinfo {title} {Size-dependent surface free energy and
  tolman-corrected droplet nucleation of tip4p/2005 water},}\ }\href {\doibase
  10.1021/jz402226p} {\bibfield  {journal} {\bibinfo  {journal} {The Journal of
  Physical Chemistry Letters}\ }\textbf {\bibinfo {volume} {4}},\ \bibinfo
  {pages} {4267--4272} (\bibinfo {year} {2013})},\ \bibinfo {note} {pMID:
  26296177}\BibitemShut {NoStop}%
\bibitem [{\citenamefont {Joswiak}\ \emph {et~al.}(2016)\citenamefont
  {Joswiak}, \citenamefont {Do}, \citenamefont {Doherty},\ and\ \citenamefont
  {Peters}}]{Joswiak2016}%
  \BibitemOpen
  \bibfield  {author} {\bibinfo {author} {\bibfnamefont {M.~N.}\ \bibnamefont
  {Joswiak}}, \bibinfo {author} {\bibfnamefont {R.}~\bibnamefont {Do}},
  \bibinfo {author} {\bibfnamefont {M.~F.}\ \bibnamefont {Doherty}}, \ and\
  \bibinfo {author} {\bibfnamefont {B.}~\bibnamefont {Peters}},\ }\bibfield
  {title} {\enquote {\bibinfo {title} {Energetic and entropic components of the
  tolman length for mw and tip4p/2005 water nanodroplets},}\ }\href {\doibase
  10.1063/1.4967875} {\bibfield  {journal} {\bibinfo  {journal} {Journal of
  Chemical Physics}\ }\textbf {\bibinfo {volume} {145}},\ \bibinfo {pages}
  {204703} (\bibinfo {year} {2016})}\BibitemShut {NoStop}%
\bibitem [{\citenamefont {Kandu\v{c}}(2017)}]{Kanduc2017}%
  \BibitemOpen
  \bibfield  {author} {\bibinfo {author} {\bibfnamefont {M.}~\bibnamefont
  {Kandu\v{c}}},\ }\bibfield  {title} {\enquote {\bibinfo {title} {Going beyond
  the standard line tension: Size-dependent contact angles of water
  nanodroplets},}\ }\href {\doibase 10.1063/1.4990741} {\bibfield  {journal}
  {\bibinfo  {journal} {The Journal of Chemical Physics}\ }\textbf {\bibinfo
  {volume} {147}},\ \bibinfo {pages} {174701} (\bibinfo {year}
  {2017})}\BibitemShut {NoStop}%
\bibitem [{\citenamefont {Leong}\ and\ \citenamefont {Wang}(2018)}]{Leong2018}%
  \BibitemOpen
  \bibfield  {author} {\bibinfo {author} {\bibfnamefont {K.-Y.}\ \bibnamefont
  {Leong}}\ and\ \bibinfo {author} {\bibfnamefont {F.}~\bibnamefont {Wang}},\
  }\bibfield  {title} {\enquote {\bibinfo {title} {A molecular dynamics
  investigation of the surface tension of water nanodroplets and a new
  technique for local pressure determination through density correlation},}\
  }\href {\doibase 10.1063/1.5004985} {\bibfield  {journal} {\bibinfo
  {journal} {The Journal of Chemical Physics}\ }\textbf {\bibinfo {volume}
  {148}},\ \bibinfo {pages} {144503} (\bibinfo {year} {2018})}\BibitemShut
  {NoStop}%
\bibitem [{\citenamefont {Fisher}\ and\ \citenamefont
  {Wortis}(1984)}]{Fisher1984}%
  \BibitemOpen
  \bibfield  {author} {\bibinfo {author} {\bibfnamefont {M.~P.~A.}\
  \bibnamefont {Fisher}}\ and\ \bibinfo {author} {\bibfnamefont
  {M.}~\bibnamefont {Wortis}},\ }\bibfield  {title} {\enquote {\bibinfo {title}
  {Curvature corrections to the surface tension of fluid drops: Landau theory
  and a scaling hypothesis},}\ }\href {\doibase 10.1103/PhysRevB.29.6252}
  {\bibfield  {journal} {\bibinfo  {journal} {Physical Review B}\ }\textbf
  {\bibinfo {volume} {29}},\ \bibinfo {pages} {6252--6260} (\bibinfo {year}
  {1984})}\BibitemShut {NoStop}%
\bibitem [{\citenamefont {Blokhuis}\ and\ \citenamefont
  {Bedeaux}(1993)}]{Blokhuis1993}%
  \BibitemOpen
  \bibfield  {author} {\bibinfo {author} {\bibfnamefont {E.~M.}\ \bibnamefont
  {Blokhuis}}\ and\ \bibinfo {author} {\bibfnamefont {D.}~\bibnamefont
  {Bedeaux}},\ }\bibfield  {title} {\enquote {\bibinfo {title} {Van der waals
  theory of curved surfaces},}\ }\href {\doibase 10.1080/00268979300102581}
  {\bibfield  {journal} {\bibinfo  {journal} {Molecular Physics}\ }\textbf
  {\bibinfo {volume} {80}},\ \bibinfo {pages} {705--720} (\bibinfo {year}
  {1993})}\BibitemShut {NoStop}%
\bibitem [{\citenamefont {van Giessen}, \citenamefont {Blokhuis},\ and\
  \citenamefont {Bukman}(1998)}]{Giessen1998}%
  \BibitemOpen
  \bibfield  {author} {\bibinfo {author} {\bibfnamefont {A.~E.}\ \bibnamefont
  {van Giessen}}, \bibinfo {author} {\bibfnamefont {E.~M.}\ \bibnamefont
  {Blokhuis}}, \ and\ \bibinfo {author} {\bibfnamefont {D.~J.}\ \bibnamefont
  {Bukman}},\ }\bibfield  {title} {\enquote {\bibinfo {title} {Mean field
  curvature corrections to the surface tension},}\ }\href {\doibase
  10.1063/1.475477} {\bibfield  {journal} {\bibinfo  {journal} {The Journal of
  Chemical Physics}\ }\textbf {\bibinfo {volume} {108}},\ \bibinfo {pages}
  {1148--1156} (\bibinfo {year} {1998})}\BibitemShut {NoStop}%
\bibitem [{\citenamefont {Barrett}(2009)}]{Barrett2009}%
  \BibitemOpen
  \bibfield  {author} {\bibinfo {author} {\bibfnamefont {J.~C.}\ \bibnamefont
  {Barrett}},\ }\bibfield  {title} {\enquote {\bibinfo {title} {On the
  thermodynamic expansion of the nucleation free-energy barrier},}\ }\href
  {\doibase 10.1063/1.3173196} {\bibfield  {journal} {\bibinfo  {journal} {The
  Journal of Chemical Physics}\ }\textbf {\bibinfo {volume} {131}},\ \bibinfo
  {pages} {084711} (\bibinfo {year} {2009})}\BibitemShut {NoStop}%
\bibitem [{\citenamefont {Rehner}\ and\ \citenamefont
  {Gross}(2018{\natexlab{b}})}]{Rehner2018a}%
  \BibitemOpen
  \bibfield  {author} {\bibinfo {author} {\bibfnamefont {P.}~\bibnamefont
  {Rehner}}\ and\ \bibinfo {author} {\bibfnamefont {J.}~\bibnamefont {Gross}},\
  }\bibfield  {title} {\enquote {\bibinfo {title} {Predictive density gradient
  theory based on nonlocal density functional theory},}\ }\href {\doibase
  10.1103/PhysRevE.98.063312} {\bibfield  {journal} {\bibinfo  {journal}
  {Physical Review E}\ }\textbf {\bibinfo {volume} {98}},\ \bibinfo {pages}
  {063312} (\bibinfo {year} {2018}{\natexlab{b}})}\BibitemShut {NoStop}%
\bibitem [{\citenamefont {Koenig}(1950)}]{Koenig1950}%
  \BibitemOpen
  \bibfield  {author} {\bibinfo {author} {\bibfnamefont {F.~O.}\ \bibnamefont
  {Koenig}},\ }\bibfield  {title} {\enquote {\bibinfo {title} {On the
  thermodynamic relation between surface tension and curvature},}\ }\href
  {\doibase 10.1063/1.1747660} {\bibfield  {journal} {\bibinfo  {journal} {The
  Journal of Chemical Physics}\ }\textbf {\bibinfo {volume} {18}},\ \bibinfo
  {pages} {449--459} (\bibinfo {year} {1950})}\BibitemShut {NoStop}%
\bibitem [{\citenamefont {Rosenfeld}(1989)}]{Rosenfeld1989}%
  \BibitemOpen
  \bibfield  {author} {\bibinfo {author} {\bibfnamefont {Y.}~\bibnamefont
  {Rosenfeld}},\ }\bibfield  {title} {\enquote {\bibinfo {title} {Free-energy
  model for the inhomogeneous hard-sphere fluid mixture and density-functional
  theory of freezing},}\ }\href {\doibase 10.1103/PhysRevLett.63.980}
  {\bibfield  {journal} {\bibinfo  {journal} {Physical Review Letters}\
  }\textbf {\bibinfo {volume} {63}},\ \bibinfo {pages} {980--983} (\bibinfo
  {year} {1989})}\BibitemShut {NoStop}%
\bibitem [{\citenamefont {Tarazona}(1985)}]{Tarazona1985}%
  \BibitemOpen
  \bibfield  {author} {\bibinfo {author} {\bibfnamefont {P.}~\bibnamefont
  {Tarazona}},\ }\bibfield  {title} {\enquote {\bibinfo {title} {Free-energy
  density functional for hard spheres},}\ }\href {\doibase
  10.1103/PhysRevA.31.2672} {\bibfield  {journal} {\bibinfo  {journal}
  {Physical Review A}\ }\textbf {\bibinfo {volume} {31}},\ \bibinfo {pages}
  {2672--2679} (\bibinfo {year} {1985})}\BibitemShut {NoStop}%
\bibitem [{\citenamefont {Mairhofer}\ and\ \citenamefont
  {Gross}(2017{\natexlab{a}})}]{Mairhofer2017}%
  \BibitemOpen
  \bibfield  {author} {\bibinfo {author} {\bibfnamefont {J.}~\bibnamefont
  {Mairhofer}}\ and\ \bibinfo {author} {\bibfnamefont {J.}~\bibnamefont
  {Gross}},\ }\bibfield  {title} {\enquote {\bibinfo {title} {Numerical aspects
  of classical density functional theory for one-dimensional vapor-liquid
  interfaces},}\ }\href {\doibase http://doi.org/10.1016/j.fluid.2017.03.023}
  {\bibfield  {journal} {\bibinfo  {journal} {Fluid Phase Equilibria}\ }\textbf
  {\bibinfo {volume} {444}},\ \bibinfo {pages} {1--12} (\bibinfo {year}
  {2017}{\natexlab{a}})}\BibitemShut {NoStop}%
\bibitem [{\citenamefont {Anderson}(1965)}]{Anderson1965}%
  \BibitemOpen
  \bibfield  {author} {\bibinfo {author} {\bibfnamefont {D.~G.}\ \bibnamefont
  {Anderson}},\ }\bibfield  {title} {\enquote {\bibinfo {title} {Iterative
  procedures for nonlinear integral equations},}\ }\href {\doibase
  10.1145/321296.321305} {\bibfield  {journal} {\bibinfo  {journal} {Journal of
  the ACM}\ }\textbf {\bibinfo {volume} {12}},\ \bibinfo {pages} {547--560}
  (\bibinfo {year} {1965})}\BibitemShut {NoStop}%
\bibitem [{\citenamefont {Miqueu}\ \emph {et~al.}(2004)\citenamefont {Miqueu},
  \citenamefont {Mendiboure}, \citenamefont {Graciaa},\ and\ \citenamefont
  {Lachaise}}]{Miqueu2004}%
  \BibitemOpen
  \bibfield  {author} {\bibinfo {author} {\bibfnamefont {C.}~\bibnamefont
  {Miqueu}}, \bibinfo {author} {\bibfnamefont {B.}~\bibnamefont {Mendiboure}},
  \bibinfo {author} {\bibfnamefont {C.}~\bibnamefont {Graciaa}}, \ and\
  \bibinfo {author} {\bibfnamefont {J.}~\bibnamefont {Lachaise}},\ }\bibfield
  {title} {\enquote {\bibinfo {title} {Modelling of the surface tension of
  binary and ternary mixtures with the gradient theory of fluid interfaces},}\
  }\href {\doibase https://doi.org/10.1016/j.fluid.2003.12.008} {\bibfield
  {journal} {\bibinfo  {journal} {Fluid Phase Equilibria}\ }\textbf {\bibinfo
  {volume} {218}},\ \bibinfo {pages} {189 -- 203} (\bibinfo {year}
  {2004})}\BibitemShut {NoStop}%
\bibitem [{\citenamefont {Miqueu}\ \emph {et~al.}(2005)\citenamefont {Miqueu},
  \citenamefont {Mendiboure}, \citenamefont {Graciaa},\ and\ \citenamefont
  {Lachaise}}]{Miqueu2005}%
  \BibitemOpen
  \bibfield  {author} {\bibinfo {author} {\bibfnamefont {C.}~\bibnamefont
  {Miqueu}}, \bibinfo {author} {\bibfnamefont {B.}~\bibnamefont {Mendiboure}},
  \bibinfo {author} {\bibfnamefont {A.}~\bibnamefont {Graciaa}}, \ and\
  \bibinfo {author} {\bibfnamefont {J.}~\bibnamefont {Lachaise}},\ }\bibfield
  {title} {\enquote {\bibinfo {title} {Modeling of the surface tension of
  multicomponent mixtures with the gradient theory of fluid interfaces},}\
  }\href {\doibase 10.1021/ie049086l} {\bibfield  {journal} {\bibinfo
  {journal} {Industrial \& Engineering Chemistry Research}\ }\textbf {\bibinfo
  {volume} {44}},\ \bibinfo {pages} {3321--3329} (\bibinfo {year}
  {2005})}\BibitemShut {NoStop}%
\bibitem [{\citenamefont {Mairhofer}\ and\ \citenamefont
  {Gross}(2017{\natexlab{b}})}]{Mairhofer2017a}%
  \BibitemOpen
  \bibfield  {author} {\bibinfo {author} {\bibfnamefont {J.}~\bibnamefont
  {Mairhofer}}\ and\ \bibinfo {author} {\bibfnamefont {J.}~\bibnamefont
  {Gross}},\ }\bibfield  {title} {\enquote {\bibinfo {title} {Modeling of
  interfacial properties of multicomponent systems using density gradient
  theory and pcp-saft},}\ }\href {\doibase 10.1016/j.fluid.2017.02.009}
  {\bibfield  {journal} {\bibinfo  {journal} {Fluid Phase Equilibria}\ }\textbf
  {\bibinfo {volume} {439}},\ \bibinfo {pages} {31--42} (\bibinfo {year}
  {2017}{\natexlab{b}})}\BibitemShut {NoStop}%
\bibitem [{\citenamefont {Kou}, \citenamefont {Sun},\ and\ \citenamefont
  {Wang}(2015)}]{Kou2015}%
  \BibitemOpen
  \bibfield  {author} {\bibinfo {author} {\bibfnamefont {J.}~\bibnamefont
  {Kou}}, \bibinfo {author} {\bibfnamefont {S.}~\bibnamefont {Sun}}, \ and\
  \bibinfo {author} {\bibfnamefont {X.}~\bibnamefont {Wang}},\ }\bibfield
  {title} {\enquote {\bibinfo {title} {Efficient numerical methods for
  simulating surface tension of multi-component mixtures with the gradient
  theory of fluid interfaces},}\ }\href {\doibase
  https://doi.org/10.1016/j.cma.2014.10.023} {\bibfield  {journal} {\bibinfo
  {journal} {Computer Methods in Applied Mechanics and Engineering}\ }\textbf
  {\bibinfo {volume} {292}},\ \bibinfo {pages} {92 -- 106} (\bibinfo {year}
  {2015})},\ \bibinfo {note} {special Issue on Advances in Simulations of
  Subsurface Flow and Transport (Honoring Professor Mary F.
  Wheeler)}\BibitemShut {NoStop}%
\bibitem [{\citenamefont {Liang}, \citenamefont {Michelsen},\ and\
  \citenamefont {Kontogeorgis}(2016)}]{Liang2016}%
  \BibitemOpen
  \bibfield  {author} {\bibinfo {author} {\bibfnamefont {X.}~\bibnamefont
  {Liang}}, \bibinfo {author} {\bibfnamefont {M.~L.}\ \bibnamefont
  {Michelsen}}, \ and\ \bibinfo {author} {\bibfnamefont {G.~M.}\ \bibnamefont
  {Kontogeorgis}},\ }\bibfield  {title} {\enquote {\bibinfo {title} {A density
  gradient theory based method for surface tension calculations},}\ }\href
  {\doibase https://doi.org/10.1016/j.fluid.2016.06.017} {\bibfield  {journal}
  {\bibinfo  {journal} {Fluid Phase Equilibria}\ }\textbf {\bibinfo {volume}
  {428}},\ \bibinfo {pages} {153 -- 163} (\bibinfo {year} {2016})},\ \bibinfo
  {note} {theo W. de Loos Festschrift}\BibitemShut {NoStop}%
\bibitem [{\citenamefont {Gross}\ and\ \citenamefont
  {Sadowski}(2001)}]{Gross2001}%
  \BibitemOpen
  \bibfield  {author} {\bibinfo {author} {\bibfnamefont {J.}~\bibnamefont
  {Gross}}\ and\ \bibinfo {author} {\bibfnamefont {G.}~\bibnamefont
  {Sadowski}},\ }\bibfield  {title} {\enquote {\bibinfo {title}
  {Perturbed-chain saft: An equation of state based on a perturbation theory
  for chain molecules},}\ }\href {\doibase 10.1021/ie0003887} {\bibfield
  {journal} {\bibinfo  {journal} {Industrial \& Engineering Chemistry
  Research}\ }\textbf {\bibinfo {volume} {40}},\ \bibinfo {pages} {1244--1260}
  (\bibinfo {year} {2001})}\BibitemShut {NoStop}%
\bibitem [{\citenamefont {Gross}(2005)}]{Gross2005}%
  \BibitemOpen
  \bibfield  {author} {\bibinfo {author} {\bibfnamefont {J.}~\bibnamefont
  {Gross}},\ }\bibfield  {title} {\enquote {\bibinfo {title} {An
  equation-of-state contribution for polar components: Quadrupolar
  molecules},}\ }\href {\doibase 10.1002/aic.10502} {\bibfield  {journal}
  {\bibinfo  {journal} {AIChE Journal}\ }\textbf {\bibinfo {volume} {51}},\
  \bibinfo {pages} {2556--2568} (\bibinfo {year} {2005})}\BibitemShut {NoStop}%
\bibitem [{\citenamefont {Gross}\ and\ \citenamefont
  {Vrabec}(2006)}]{Gross2006}%
  \BibitemOpen
  \bibfield  {author} {\bibinfo {author} {\bibfnamefont {J.}~\bibnamefont
  {Gross}}\ and\ \bibinfo {author} {\bibfnamefont {J.}~\bibnamefont {Vrabec}},\
  }\bibfield  {title} {\enquote {\bibinfo {title} {An equation-of-state
  contribution for polar components: Dipolar molecules},}\ }\href {\doibase
  10.1002/aic.10683} {\bibfield  {journal} {\bibinfo  {journal} {AIChE
  Journal}\ }\textbf {\bibinfo {volume} {52}},\ \bibinfo {pages} {1194--1204}
  (\bibinfo {year} {2006})}\BibitemShut {NoStop}%
\bibitem [{\citenamefont {Roth}(2010)}]{Roth2010}%
  \BibitemOpen
  \bibfield  {author} {\bibinfo {author} {\bibfnamefont {R.}~\bibnamefont
  {Roth}},\ }\bibfield  {title} {\enquote {\bibinfo {title} {Fundamental
  measure theory for hard-sphere mixtures: a review},}\ }\href {\doibase
  10.1088/0953-8984/22/6/063102} {\bibfield  {journal} {\bibinfo  {journal}
  {Journal of Physics: Condensed Matter}\ }\textbf {\bibinfo {volume} {22}},\
  \bibinfo {pages} {063102} (\bibinfo {year} {2010})}\BibitemShut {NoStop}%
\bibitem [{\citenamefont {Roth}\ \emph {et~al.}(2002)\citenamefont {Roth},
  \citenamefont {Evans}, \citenamefont {Lang},\ and\ \citenamefont
  {Kahl}}]{Roth2002}%
  \BibitemOpen
  \bibfield  {author} {\bibinfo {author} {\bibfnamefont {R.}~\bibnamefont
  {Roth}}, \bibinfo {author} {\bibfnamefont {R.}~\bibnamefont {Evans}},
  \bibinfo {author} {\bibfnamefont {A.}~\bibnamefont {Lang}}, \ and\ \bibinfo
  {author} {\bibfnamefont {G.}~\bibnamefont {Kahl}},\ }\bibfield  {title}
  {\enquote {\bibinfo {title} {Fundamental measure theory for hard-sphere
  mixtures revisited: the white bear version},}\ }\href {\doibase
  10.1088/0953-8984/14/46/313} {\bibfield  {journal} {\bibinfo  {journal}
  {Journal of Physics: Condensed Matter}\ }\textbf {\bibinfo {volume} {14}},\
  \bibinfo {pages} {12063} (\bibinfo {year} {2002})}\BibitemShut {NoStop}%
\bibitem [{\citenamefont {Yu}\ and\ \citenamefont
  {Wu}(2002{\natexlab{a}})}]{Yu2002}%
  \BibitemOpen
  \bibfield  {author} {\bibinfo {author} {\bibfnamefont {Y.-X.}\ \bibnamefont
  {Yu}}\ and\ \bibinfo {author} {\bibfnamefont {J.}~\bibnamefont {Wu}},\
  }\bibfield  {title} {\enquote {\bibinfo {title} {Structures of hard-sphere
  fluids from a modified fundamental-measure theory},}\ }\href {\doibase
  10.1063/1.1520530} {\bibfield  {journal} {\bibinfo  {journal} {The Journal of
  Chemical Physics}\ }\textbf {\bibinfo {volume} {117}},\ \bibinfo {pages}
  {10156--10164} (\bibinfo {year} {2002}{\natexlab{a}})}\BibitemShut {NoStop}%
\bibitem [{\citenamefont {Kierlik}\ and\ \citenamefont
  {Rosinberg}(1990)}]{Kierlik1990}%
  \BibitemOpen
  \bibfield  {author} {\bibinfo {author} {\bibfnamefont {E.}~\bibnamefont
  {Kierlik}}\ and\ \bibinfo {author} {\bibfnamefont {M.~L.}\ \bibnamefont
  {Rosinberg}},\ }\bibfield  {title} {\enquote {\bibinfo {title} {Free-energy
  density functional for the inhomogeneous hard-sphere fluid: Application to
  interfacial adsorption},}\ }\href {\doibase 10.1103/PhysRevA.42.3382}
  {\bibfield  {journal} {\bibinfo  {journal} {Physical Review A}\ }\textbf
  {\bibinfo {volume} {42}},\ \bibinfo {pages} {3382--3387} (\bibinfo {year}
  {1990})}\BibitemShut {NoStop}%
\bibitem [{\citenamefont {Boubl{\'i}k}(1970)}]{Boublik1970}%
  \BibitemOpen
  \bibfield  {author} {\bibinfo {author} {\bibfnamefont {T.}~\bibnamefont
  {Boubl{\'i}k}},\ }\bibfield  {title} {\enquote {\bibinfo {title} {Hard-sphere
  equation of state},}\ }\href {\doibase 10.1063/1.1673824} {\bibfield
  {journal} {\bibinfo  {journal} {The Journal of Chemical Physics}\ }\textbf
  {\bibinfo {volume} {53}},\ \bibinfo {pages} {471--472} (\bibinfo {year}
  {1970})}\BibitemShut {NoStop}%
\bibitem [{\citenamefont {Mansoori}\ \emph {et~al.}(1971)\citenamefont
  {Mansoori}, \citenamefont {Carnahan}, \citenamefont {Starling},\ and\
  \citenamefont {Leland}}]{Mansoori1971}%
  \BibitemOpen
  \bibfield  {author} {\bibinfo {author} {\bibfnamefont {G.~A.}\ \bibnamefont
  {Mansoori}}, \bibinfo {author} {\bibfnamefont {N.~F.}\ \bibnamefont
  {Carnahan}}, \bibinfo {author} {\bibfnamefont {K.~E.}\ \bibnamefont
  {Starling}}, \ and\ \bibinfo {author} {\bibfnamefont {T.~W.}\ \bibnamefont
  {Leland}},\ }\bibfield  {title} {\enquote {\bibinfo {title} {Equilibrium
  thermodynamic properties of the mixture of hard spheres},}\ }\href {\doibase
  10.1063/1.1675048} {\bibfield  {journal} {\bibinfo  {journal} {The Journal of
  Chemical Physics}\ }\textbf {\bibinfo {volume} {54}},\ \bibinfo {pages}
  {1523--1525} (\bibinfo {year} {1971})}\BibitemShut {NoStop}%
\bibitem [{\citenamefont {Carnahan}\ and\ \citenamefont
  {Starling}(1969)}]{Carnahan1969}%
  \BibitemOpen
  \bibfield  {author} {\bibinfo {author} {\bibfnamefont {N.~F.}\ \bibnamefont
  {Carnahan}}\ and\ \bibinfo {author} {\bibfnamefont {K.~E.}\ \bibnamefont
  {Starling}},\ }\bibfield  {title} {\enquote {\bibinfo {title} {Equation of
  state for nonattracting rigid spheres},}\ }\href {\doibase 10.1063/1.1672048}
  {\bibfield  {journal} {\bibinfo  {journal} {The Journal of Chemical Physics}\
  }\textbf {\bibinfo {volume} {51}},\ \bibinfo {pages} {635--636} (\bibinfo
  {year} {1969})}\BibitemShut {NoStop}%
\bibitem [{\citenamefont {Tripathi}\ and\ \citenamefont
  {Chapman}(2005{\natexlab{a}})}]{Tripathi2005}%
  \BibitemOpen
  \bibfield  {author} {\bibinfo {author} {\bibfnamefont {S.}~\bibnamefont
  {Tripathi}}\ and\ \bibinfo {author} {\bibfnamefont {W.~G.}\ \bibnamefont
  {Chapman}},\ }\bibfield  {title} {\enquote {\bibinfo {title} {Microstructure
  of inhomogeneous polyatomic mixtures from a density functional formalism for
  atomic mixtures},}\ }\href {\doibase 10.1063/1.1853371} {\bibfield  {journal}
  {\bibinfo  {journal} {The Journal of Chemical Physics}\ }\textbf {\bibinfo
  {volume} {122}},\ \bibinfo {pages} {094506} (\bibinfo {year}
  {2005}{\natexlab{a}})}\BibitemShut {NoStop}%
\bibitem [{\citenamefont {Tripathi}\ and\ \citenamefont
  {Chapman}(2005{\natexlab{b}})}]{Tripathi2005a}%
  \BibitemOpen
  \bibfield  {author} {\bibinfo {author} {\bibfnamefont {S.}~\bibnamefont
  {Tripathi}}\ and\ \bibinfo {author} {\bibfnamefont {W.~G.}\ \bibnamefont
  {Chapman}},\ }\bibfield  {title} {\enquote {\bibinfo {title} {Microstructure
  and thermodynamics of inhomogeneous polymer blends and solutions},}\ }\href
  {\doibase 10.1103/PhysRevLett.94.087801} {\bibfield  {journal} {\bibinfo
  {journal} {Physical Review Letters}\ }\textbf {\bibinfo {volume} {94}},\
  \bibinfo {pages} {087801} (\bibinfo {year} {2005}{\natexlab{b}})}\BibitemShut
  {NoStop}%
\bibitem [{\citenamefont {Yu}\ and\ \citenamefont
  {Wu}(2002{\natexlab{b}})}]{Yu2002b}%
  \BibitemOpen
  \bibfield  {author} {\bibinfo {author} {\bibfnamefont {Y.-X.}\ \bibnamefont
  {Yu}}\ and\ \bibinfo {author} {\bibfnamefont {J.}~\bibnamefont {Wu}},\
  }\bibfield  {title} {\enquote {\bibinfo {title} {A fundamental-measure theory
  for inhomogeneous associating fluids},}\ }\href {\doibase 10.1063/1.1463435}
  {\bibfield  {journal} {\bibinfo  {journal} {The Journal of Chemical Physics}\
  }\textbf {\bibinfo {volume} {116}},\ \bibinfo {pages} {7094--7103} (\bibinfo
  {year} {2002}{\natexlab{b}})}\BibitemShut {NoStop}%
\bibitem [{\citenamefont {Sauer}\ and\ \citenamefont
  {Gross}(2017)}]{Sauer2017}%
  \BibitemOpen
  \bibfield  {author} {\bibinfo {author} {\bibfnamefont {E.}~\bibnamefont
  {Sauer}}\ and\ \bibinfo {author} {\bibfnamefont {J.}~\bibnamefont {Gross}},\
  }\bibfield  {title} {\enquote {\bibinfo {title} {Classical density functional
  theory for liquid–fluid interfaces and confined systems: A functional for
  the perturbed-chain polar statistical associating fluid theory equation of
  state},}\ }\href {\doibase 10.1021/acs.iecr.6b04551} {\bibfield  {journal}
  {\bibinfo  {journal} {Industrial \& Engineering Chemistry Research}\ }\textbf
  {\bibinfo {volume} {56}},\ \bibinfo {pages} {4119--4135} (\bibinfo {year}
  {2017})}\BibitemShut {NoStop}%
\bibitem [{\citenamefont {Sauer}\ \emph {et~al.}(2018)\citenamefont {Sauer},
  \citenamefont {Terzis}, \citenamefont {Theiss}, \citenamefont {Weigand},\
  and\ \citenamefont {Gross}}]{Sauer2018}%
  \BibitemOpen
  \bibfield  {author} {\bibinfo {author} {\bibfnamefont {E.}~\bibnamefont
  {Sauer}}, \bibinfo {author} {\bibfnamefont {A.}~\bibnamefont {Terzis}},
  \bibinfo {author} {\bibfnamefont {M.}~\bibnamefont {Theiss}}, \bibinfo
  {author} {\bibfnamefont {B.}~\bibnamefont {Weigand}}, \ and\ \bibinfo
  {author} {\bibfnamefont {J.}~\bibnamefont {Gross}},\ }\bibfield  {title}
  {\enquote {\bibinfo {title} {Prediction of contact angles and density
  profiles of sessile droplets using classical density functional theory based
  on the pcp-saft equation of state},}\ }\href {\doibase
  10.1021/acs.langmuir.8b01985} {\bibfield  {journal} {\bibinfo  {journal}
  {Langmuir}\ }\textbf {\bibinfo {volume} {34}},\ \bibinfo {pages}
  {12519--12531} (\bibinfo {year} {2018})},\ \bibinfo {note} {pMID:
  30247038}\BibitemShut {NoStop}%
\bibitem [{\citenamefont {Sauer}\ and\ \citenamefont
  {Gross}(2019)}]{Sauer2019}%
  \BibitemOpen
  \bibfield  {author} {\bibinfo {author} {\bibfnamefont {E.}~\bibnamefont
  {Sauer}}\ and\ \bibinfo {author} {\bibfnamefont {J.}~\bibnamefont {Gross}},\
  }\bibfield  {title} {\enquote {\bibinfo {title} {Prediction of adsorption
  isotherms and selectivities: Comparison between classical density functional
  theory based on the perturbed-chain statistical associating fluid theory
  equation of state and ideal adsorbed solution theory},}\ }\href {\doibase
  10.1021/acs.langmuir.9b02378} {\bibfield  {journal} {\bibinfo  {journal}
  {Langmuir}\ }\textbf {\bibinfo {volume} {35}},\ \bibinfo {pages}
  {11690--11701} (\bibinfo {year} {2019})},\ \bibinfo {note} {pMID:
  31403314}\BibitemShut {NoStop}%
\bibitem [{\citenamefont {Klink}\ and\ \citenamefont
  {Gross}(2014)}]{Klink2014}%
  \BibitemOpen
  \bibfield  {author} {\bibinfo {author} {\bibfnamefont {C.}~\bibnamefont
  {Klink}}\ and\ \bibinfo {author} {\bibfnamefont {J.}~\bibnamefont {Gross}},\
  }\bibfield  {title} {\enquote {\bibinfo {title} {A density functional theory
  for vapor-liquid interfaces of mixtures using the perturbed-chain polar
  statistical associating fluid theory equation of state},}\ }\href {\doibase
  10.1021/ie4029895} {\bibfield  {journal} {\bibinfo  {journal} {Industrial \&
  Engineering Chemistry Research}\ }\textbf {\bibinfo {volume} {53}},\ \bibinfo
  {pages} {6169--6178} (\bibinfo {year} {2014})}\BibitemShut {NoStop}%
\bibitem [{\citenamefont {Mulero}, \citenamefont {Cachadiña},\ and\
  \citenamefont {Parra}(2012)}]{Mulero2012}%
  \BibitemOpen
  \bibfield  {author} {\bibinfo {author} {\bibfnamefont {A.}~\bibnamefont
  {Mulero}}, \bibinfo {author} {\bibfnamefont {I.}~\bibnamefont {Cachadiña}},
  \ and\ \bibinfo {author} {\bibfnamefont {M.~I.}\ \bibnamefont {Parra}},\
  }\bibfield  {title} {\enquote {\bibinfo {title} {Recommended correlations for
  the surface tension of common fluids},}\ }\href {\doibase 10.1063/1.4768782}
  {\bibfield  {journal} {\bibinfo  {journal} {Journal of Physical and Chemical
  Reference Data}\ }\textbf {\bibinfo {volume} {41}},\ \bibinfo {pages}
  {043105} (\bibinfo {year} {2012})}\BibitemShut {NoStop}%
\bibitem [{\citenamefont {Mairhofer}\ and\ \citenamefont
  {Gross}(2018)}]{Mairhofer2018}%
  \BibitemOpen
  \bibfield  {author} {\bibinfo {author} {\bibfnamefont {J.}~\bibnamefont
  {Mairhofer}}\ and\ \bibinfo {author} {\bibfnamefont {J.}~\bibnamefont
  {Gross}},\ }\bibfield  {title} {\enquote {\bibinfo {title} {Modeling
  properties of the one-dimensional vapor-liquid interface: Application of
  classical density functional and density gradient theory},}\ }\href {\doibase
  https://doi.org/10.1016/j.fluid.2017.11.032} {\bibfield  {journal} {\bibinfo
  {journal} {Fluid Phase Equilibria}\ }\textbf {\bibinfo {volume} {458}},\
  \bibinfo {pages} {243 -- 252} (\bibinfo {year} {2018})}\BibitemShut {NoStop}%
\bibitem [{\citenamefont {Lemmon}, \citenamefont {McLinden},\ and\
  \citenamefont {Friend}()}]{Lemmon}%
  \BibitemOpen
  \bibfield  {author} {\bibinfo {author} {\bibfnamefont {E.~W.}\ \bibnamefont
  {Lemmon}}, \bibinfo {author} {\bibfnamefont {M.~O.}\ \bibnamefont
  {McLinden}}, \ and\ \bibinfo {author} {\bibfnamefont {D.~G.}\ \bibnamefont
  {Friend}},\ }\enquote {\bibinfo {title} {Nist chemistry webbook, nist
  standard reference database number 69},}\ \ (\bibinfo  {publisher} {National
  Institute of Standards and Technology},\ \bibinfo {address} {Gaithersburg MD,
  20899})\ Chap.\ \bibinfo {chapter} {Thermophysical Properties of Fluid
  Systems},\ \bibinfo {note} {, http://webbook.nist.gov, (retrieved July 15,
  2016)}\BibitemShut {NoStop}%
\bibitem [{\citenamefont {Huang}\ and\ \citenamefont
  {Radosz}(1990)}]{Huang1990}%
  \BibitemOpen
  \bibfield  {author} {\bibinfo {author} {\bibfnamefont {S.~H.}\ \bibnamefont
  {Huang}}\ and\ \bibinfo {author} {\bibfnamefont {M.}~\bibnamefont {Radosz}},\
  }\bibfield  {title} {\enquote {\bibinfo {title} {Equation of state for small,
  large, polydisperse, and associating molecules},}\ }\href {\doibase
  10.1021/ie00107a014} {\bibfield  {journal} {\bibinfo  {journal} {Industrial
  \& Engineering Chemistry Research}\ }\textbf {\bibinfo {volume} {29}},\
  \bibinfo {pages} {2284--2294} (\bibinfo {year} {1990})}\BibitemShut {NoStop}%
\bibitem [{\citenamefont {Kontogeorgis}\ \emph {et~al.}(1996)\citenamefont
  {Kontogeorgis}, \citenamefont {Voutsas}, \citenamefont {Yakoumis},\ and\
  \citenamefont {Tassios}}]{Kontogeorgis1996}%
  \BibitemOpen
  \bibfield  {author} {\bibinfo {author} {\bibfnamefont {G.~M.}\ \bibnamefont
  {Kontogeorgis}}, \bibinfo {author} {\bibfnamefont {E.~C.}\ \bibnamefont
  {Voutsas}}, \bibinfo {author} {\bibfnamefont {I.~V.}\ \bibnamefont
  {Yakoumis}}, \ and\ \bibinfo {author} {\bibfnamefont {D.~P.}\ \bibnamefont
  {Tassios}},\ }\bibfield  {title} {\enquote {\bibinfo {title} {An equation of
  state for associating fluids},}\ }\href {\doibase 10.1021/ie9600203}
  {\bibfield  {journal} {\bibinfo  {journal} {Industrial \& Engineering
  Chemistry Research}\ }\textbf {\bibinfo {volume} {35}},\ \bibinfo {pages}
  {4310--4318} (\bibinfo {year} {1996})}\BibitemShut {NoStop}%
\bibitem [{\citenamefont {Fike}\ and\ \citenamefont {Alonso}(2011)}]{Fike2011}%
  \BibitemOpen
  \bibfield  {author} {\bibinfo {author} {\bibfnamefont {J.}~\bibnamefont
  {Fike}}\ and\ \bibinfo {author} {\bibfnamefont {J.}~\bibnamefont {Alonso}},\
  }\enquote {\bibinfo {title} {The development of hyper-dual numbers for exact
  second-derivative calculations},}\ in\ \href {\doibase 10.2514/6.2011-886}
  {\emph {\bibinfo {booktitle} {49th AIAA Aerospace Sciences Meeting including
  the New Horizons Forum and Aerospace Exposition}}}\ (\bibinfo {year}
  {2011})\BibitemShut {NoStop}%
\bibitem [{\citenamefont {Diewald}\ \emph {et~al.}(2018)\citenamefont
  {Diewald}, \citenamefont {Heier}, \citenamefont {Horsch}, \citenamefont
  {Kuhn}, \citenamefont {Langenbach}, \citenamefont {Hasse},\ and\
  \citenamefont {M\"uller}}]{Diewald2018}%
  \BibitemOpen
  \bibfield  {author} {\bibinfo {author} {\bibfnamefont {F.}~\bibnamefont
  {Diewald}}, \bibinfo {author} {\bibfnamefont {M.}~\bibnamefont {Heier}},
  \bibinfo {author} {\bibfnamefont {M.}~\bibnamefont {Horsch}}, \bibinfo
  {author} {\bibfnamefont {C.}~\bibnamefont {Kuhn}}, \bibinfo {author}
  {\bibfnamefont {K.}~\bibnamefont {Langenbach}}, \bibinfo {author}
  {\bibfnamefont {H.}~\bibnamefont {Hasse}}, \ and\ \bibinfo {author}
  {\bibfnamefont {R.}~\bibnamefont {M\"uller}},\ }\bibfield  {title} {\enquote
  {\bibinfo {title} {Three-dimensional phase field modeling of inhomogeneous
  gas-liquid systems using the pets equation of state},}\ }\href {\doibase
  10.1063/1.5035495} {\bibfield  {journal} {\bibinfo  {journal} {The Journal of
  Chemical Physics}\ }\textbf {\bibinfo {volume} {149}},\ \bibinfo {pages}
  {064701} (\bibinfo {year} {2018})}\BibitemShut {NoStop}%
\end{thebibliography}%
%\bibliography{literature_CE.bib}

\end{document}